\documentclass[fleqn,usenatbib]{mnras}

\usepackage{newtxtext,newtxmath}

\usepackage[T1]{fontenc}

\DeclareRobustCommand{\VAN}[3]{#2}
\let\VANthebibliography\thebibliography
\def\thebibliography{\DeclareRobustCommand{\VAN}[3]{##3}\VANthebibliography}

\usepackage{graphicx}	
\usepackage{amsmath}	
\usepackage{amssymb}	
\usepackage{xspace}
\usepackage{anyfontsize}
\usepackage{booktabs} 
\usepackage{multirow} 
\usepackage{lscape}
\usepackage{enumitem}
\usepackage{varwidth}
\usepackage{aas_macros} 
\usepackage{lineno}
\usepackage{float}

\newcommand{\redmapper}{redMaPPer\xspace}
\newcommand{\erosita}{\textit{eROSITA}\xspace}
\newcommand{\planck}{\textit{Planck}\xspace}

\newcommand{\lcdm}{$\Lambda\text{CDM}$\xspace}

\DeclareMathOperator*{\lstar}{\textit{L}^*\xspace} 
\DeclareMathOperator{\sigmacl}{\sigma_{\text{cl}}}
\DeclareMathOperator*{\pmem}{\textit{p}_{\text{mem}}}

\DeclareMathOperator{\zlambda}{\textit{z}_{\lambda}\xspace}

\DeclareMathOperator{\lambdaobs}{\lambda\xspace}
\DeclareMathOperator{\lambdatrue}{\lambda_{\text{spec}}\xspace}

\DeclareMathOperator*{\pspec}{\textit{p}_{\mathrm{spec}}}

\DeclareMathOperator*{\pred}{\textit{p}_{\mathrm{red}}}

\DeclareMathOperator*{\fproj}{\textit{f}_{\text{proj}}}

\DeclareMathOperator{\deltaznormed}{\Delta z / (1+z)\xspace}

\defcitealias{Myles2021}{M21}

\title[DESI Characterization of \redmapper{} Galaxy Clusters]{Spectroscopic Characterization of \redmapper{} Galaxy Clusters with DESI}

\author[J. Myles et al.]{
\parbox{\textwidth}{
\Large{J.~Myles,$^{1}$ \thanks{E-mail: jmyles@princeton.edu}
D.~Gruen,$^{2,3}$
T.~Jeltema,$^{4}$
A.~Mantz,$^{5,6}$
S.~Allen,$^{5,6}$
S.~Fu,$^{6}$
A.~Kremin,$^{7}$
J.~Aguilar,$^{7}$
S.~Ahlen,$^{8}$
D.~Bianchi,$^{9,10}$
D.~Brooks,$^{11}$
F.~J.~Castander,$^{12,13}$
T.~Claybaugh,$^{7}$
A.~de la Macorra,$^{14}$
Arjun~Dey,$^{15}$
P.~Doel,$^{11}$
S.~Ferraro,$^{7,16}$
J.~E.~Forero-Romero,$^{17,18}$
E.~Gaztañaga,$^{12,19,13}$
S.~Gontcho A Gontcho,$^{7,20}$
G.~Gutierrez,$^{21}$
K.~Honscheid,$^{22,23,24}$
M.~Ishak,$^{25}$
R.~Kehoe,$^{26}$
D.~Kirkby,$^{27}$
T.~Kisner,$^{7}$
O.~Lahav,$^{11}$
M.~Landriau,$^{7}$
L.~Le~Guillou,$^{28}$
M.~Manera,$^{29,30}$
A.~Meisner,$^{15}$
R.~Miquel,$^{31,30}$
J.~Moustakas,$^{32}$
S.~Nadathur,$^{19}$
J.~ A.~Newman,$^{33}$
N.~Palanque-Delabrouille,$^{34,7}$
F.~Prada,$^{35}$
I.~P\'erez-R\`afols,$^{36}$
G.~Rossi,$^{37}$
E.~Sanchez,$^{38}$
D.~Schlegel,$^{7}$
M.~Schubnell,$^{39,40}$
J.~Silber,$^{7}$
D.~Sprayberry,$^{15}$
G.~Tarl\'{e},$^{40}$
B.~A.~Weaver,$^{15}$
and R.~Zhou$^{7}$
\begin{center} (DESI Collaboration) \end{center}
}
\vspace{0.4cm}
\parbox{\textwidth}{ \small
\textit{The authors' affiliations are shown at the end of this paper.}}}}

\date{Accepted 2025 October 21. Received 2025 October 21; in original form 2025 June 5}

\pubyear{2025}

\begin{document}
\label{firstpage}
\pagerange{\pageref{firstpage}--\pageref{lastpage}}
\maketitle

\begin{abstract}
Optical galaxy cluster identification algorithms such as \redmapper{} promise to enable an array of astrophysical and cosmological studies, but suffer from biases whereby galaxies in front of and behind a galaxy cluster are mistakenly associated with the primary cluster halo. These projection effects caused by irreducible photometric redshift uncertainty must be quantified to facilitate the use of optical cluster catalogues. We present measurements of galaxy cluster projection effects and velocity dispersion using spectroscopy from the Dark Energy Spectroscopic Instrument (DESI). Our findings are as follows: we confirm that the fraction of \redmapper{} putative member galaxies mistakenly associated with cluster haloes is richness dependent, being more than twice as large at low richness than high richness; we present the first spectroscopic evidence of an increase in projection effects with increasing redshift, by as much as 25 per cent from $z\sim0.1$ to $z\sim0.2$; moreover, we find qualitative evidence for luminosity dependence in projection effects, with fainter galaxies being more commonly far behind clusters than their bright counterparts; finally we fit the scaling relation between measured mean spectroscopic richness and velocity dispersion, finding an implied linear scaling between spectroscopic richness and halo mass. We discuss further directions for the application of spectroscopic datasets to improve use of optically selected clusters to test cosmological models.
\end{abstract}





\section{Introduction}

Observations of galaxy clusters enable studies addressing a range of open problems in astrophysics and cosmology. The flagship cosmological probe using galaxy clusters is the observed number density as a function of cluster mass and redshift, which has provided competitive constraints on the mean matter density of the Universe $\Omega_{\rm{m}}$, the amplitude of matter density fluctuation $\sigma_8$, the dark energy density $\Omega_{\mathrm{DE}}$, and the dark energy equation-of-state parameter $w$ \citep[][and references therein]{Allen1103.4829}. Moreover, cluster abundance provides information on deviations of gravity from general relativity and on models of inflation \citep{Cataneo2018b, Heneka2018}. In combination with observations of the Cosmic Microwave Background (CMB), clusters have delivered competitive constraints on the parameters of the standard cosmological model \citep[e.g.,][]{WTGIV, Planck2018a, Bocquet2019, Costanzi2019, to2020a, des_y3_clusters, erosita_cosmo}. 
Among probes of the large-scale structure of the Universe, galaxy clusters stand out because they can be detected across a broad wavelength range, with each observable offering advantages complementary to its counterparts. The primary challenge in cluster cosmology is measuring cluster mass, which is the key link between cluster abundance and the halo mass function predicted by theory.
Each cluster observable lies at a distinct point in the space of trade-offs between accuracy and precision on mass at a given wavelength as well as the purity, completeness, and observational feasibility of samples identified via its data. 

The original data used to identify clusters of galaxies -- optical imaging -- has the distinct advantage of enabling readily detecting clusters at all masses \citep{Shapley1933, Abell1957}. Moreover, cluster samples determined from optical imaging are produced from the same imaging datasets used for a broad range of cosmological analyses, enabling self-consistency tests and combination of constraints \citep[e.g.,][]{to2020}. Additionally, optical imaging provides cluster photometric redshifts ($\sigma_z \sim 0.01$) and weak gravitational lensing measurements to calibrate mean cluster masses \citep{redmapper1, McClintock2019}. Optical cluster samples are especially valuable for measuring the clustering of galaxy clusters because of their sensitivity to the lowest mass clusters \citep{mana2013, to2020, to2020a, Park2023, Sunayama2024}.

For optically selected cluster samples, the observed property commonly used as a mass proxy is \textit{richness}, a (typically weighted) count of cluster galaxies \citep{Zwicky1961, redmapper1}. The \textbf{red}-sequence \textbf{Ma}tched-filter \textbf{P}robabilistic \textbf{Per}colation (\redmapper{}) algorithm, a leading algorithm for detecting and characterizing clusters from optical imaging datasets, leverages the red-sequence of galaxies to infer cluster membership and photometric redshifts. By defining \redmapper{} richness ($\lambda$) as the probability-weighted count of \textit{bright red} cluster galaxies, rather than all detected galaxies, the algorithm is able to reduce otherwise prohibitively high scatter in the mass-richness relation \citep{redmapper2}. Inevitably, however, all optical imaging cluster finders suffer from some degree of mis-classification of galaxies along the line-of-sight \citep{Lucey1983, redmapper4, Sohn1712.00872, Costanzi1807.07072, Myles2021, GrandisMohr2021, Lee2024, Grandis2025}.

This problem of projection effects introduces multiple issues of bias and scatter in the mass-observable relation for optical clusters, with such effects being correlated. First, richness is generically boosted by galaxies along the line of sight, with the boost being richness-dependent \citep[hereafter M21]{Myles2021}. This has the effect of increasing scatter in the mass-observable relation as objects from a larger range of true richness can be scattered into a bin of observed richness due to galaxies in projection. Second, the mass associated with putative cluster galaxies (i.e., galaxies identified by a cluster finding algorithm as members) which are in fact in projection contribute to the observed cluster lensing signal. Third, haloes which are in filaments aligned along the line of sight are preferentially selected by the algorithm as cluster detections, leading to boosts of the clustering and lensing of such detections on large angular scales \citep{Sunayama2020, desy1, Wu2022, Zhou2024}. Fourth, triaxial cluster haloes with a major axis aligned along the line of sight are similarly preferentially selected by the cluster finding algorithm and have distinct (boosted) lensing signals \citep{Zhang2023}. These problems compound the lensing measurement uncertainties such as the unknown `intrinsic shape' of lensed galaxies (i.e. the notional image in the absence of the effect of lensing) and the fact that lensing fundamentally measures the projected mass distribution along the entire line of sight to a galaxy cluster of interest \citep{Becker2011}. Correcting these problems requires characterizing the incidence of projection effects and the relationship between `true' richness (i.e. the three-dimensional galaxy count corresponding to richness in the absence of photometric redshift uncertainty) and observed richness. 

Studies constraining cosmology with optically selected cluster samples illustrate the importance of quantifying projection effects. The Dark Energy Survey Year 1 (DES Y1) analysis using optically selected clusters with the mass-richness relation informed by weak lensing is a prime example: this analysis identified richness-dependent systematic effects on cluster observables as the key to understanding anomalously discordant best-fitting parameters of the cosmological model \citep[characterized by, e.g., $\Omega_{\rm{m}} = 0.179^{+0.031}_{-0.038}$]{desy1}. Subsequent studies have mitigated this discrepancy in cosmological parameters  by removing small-scale lensing information or by accounting for anisotropic boosts in lensing due to projection effects \citep{to2020, Sunayama2024, Salcedo2024, des_y3_clusters}. These results motivate robust constraints on projection effects so that optically selected cluster samples may be fully utilized. 

Measurements of projection effects in simulations are in general affected by differences between the simulations and the real universe. In an idealized case, one could study projection effects using hydrodynamical simulations in which the formation and evolution of galaxies is modeled realistically and simulated over large volumes. This approach would also require realistically forward-modeled photometric observables to apply optical cluster finding algorithms to the simulated data. Given that such an approach is infeasible, we turn in this work to empirical methods. 

There are two leading observational techniques to quantify projection effects: X-ray imaging and optical galaxy spectroscopy. X-ray imaging is suitable due to the dependence of X-ray luminosity on the square of the density of the intracluster medium. Spectroscopy of cluster galaxies is suitable because the uncertainty of spectroscopic redshifts ($\lesssim 0.001$) enables high-fidelity assessment of cluster membership. Spectroscopy of cluster galaxies is an underutilized approach for characterizing galaxy clusters given its unique combination of relatively high sensitivity to low mass together with its \textit{relative} feasibility at higher redshifts compared to X-ray observations.  Unlike optical imaging, spectroscopy enables distinguishing cluster galaxies from those in projection, addressing the primary problem associated with optically selected clusters. Moreover, cluster spectroscopy, like its ICM-based counterparts, constrains a dynamical measure of cluster mass via the velocity dispersion of cluster galaxies. The era of massively multiplexed Stage-IV spectrographs [e.g., DESI, the Prime Focus Spectrograph (PFS), and the 4-meter Multi-Object Spectroscopic Telescope (4MOST)] will yield large numbers (tens of millions) of galaxy redshifts \citep{Snowmass2013.Levi,DESI2016a.Science,DESI2022.KP1.Instr, 4most, pfs}. These datasets enable measuring stacked cluster velocity dispersion and calibrating projection effects. Where survey data become sparse, targeted observations can serve a calibration role analogous to the role X-ray luminosity has played in optical cluster cosmology for decades \citep[e.g.,][]{WTGV}. 
 
The study by \citetalias{Myles2021} demonstrated a first use of representative survey spectroscopy (SDSS) to investigate projection effects and velocity dispersion in \redmapper{} clusters at $z\sim0.1$ for galaxies brighter than $0.55\lstar$. In this paper, we report on the use of Dark Energy Spectroscopic Instrument \citep[DESI]{DESI2016b.Instr,FocalPlane.Silber.2023,Corrector.Miller.2023,FiberSystem.Poppett.2024,DESI2023a.KP1.SV,DESI2023b.KP1.EDR} data to conduct measurements of projection effects and cluster velocity dispersion at the lowest galaxy luminosities used by \redmapper{} ($L\geq 0.2\lstar$ at $z\sim0.1$) as well as to higher redshift than previously possible ($z\sim0.2$). By making use of the substantially larger and fainter dataset available in DESI compared to SDSS \citep[e.g.,][]{DESI_DR1_2025}, we characterize optical cluster projection effects in greater detail. We compare our empirical results to the literature and comment on how our findings relate to recent optical cluster cosmology results. This paper represents a first step towards conducting an optical galaxy cluster cosmology analysis that makes full use of the data collected by DESI. 

\begin{figure*}
    \includegraphics[width=0.9\textwidth]{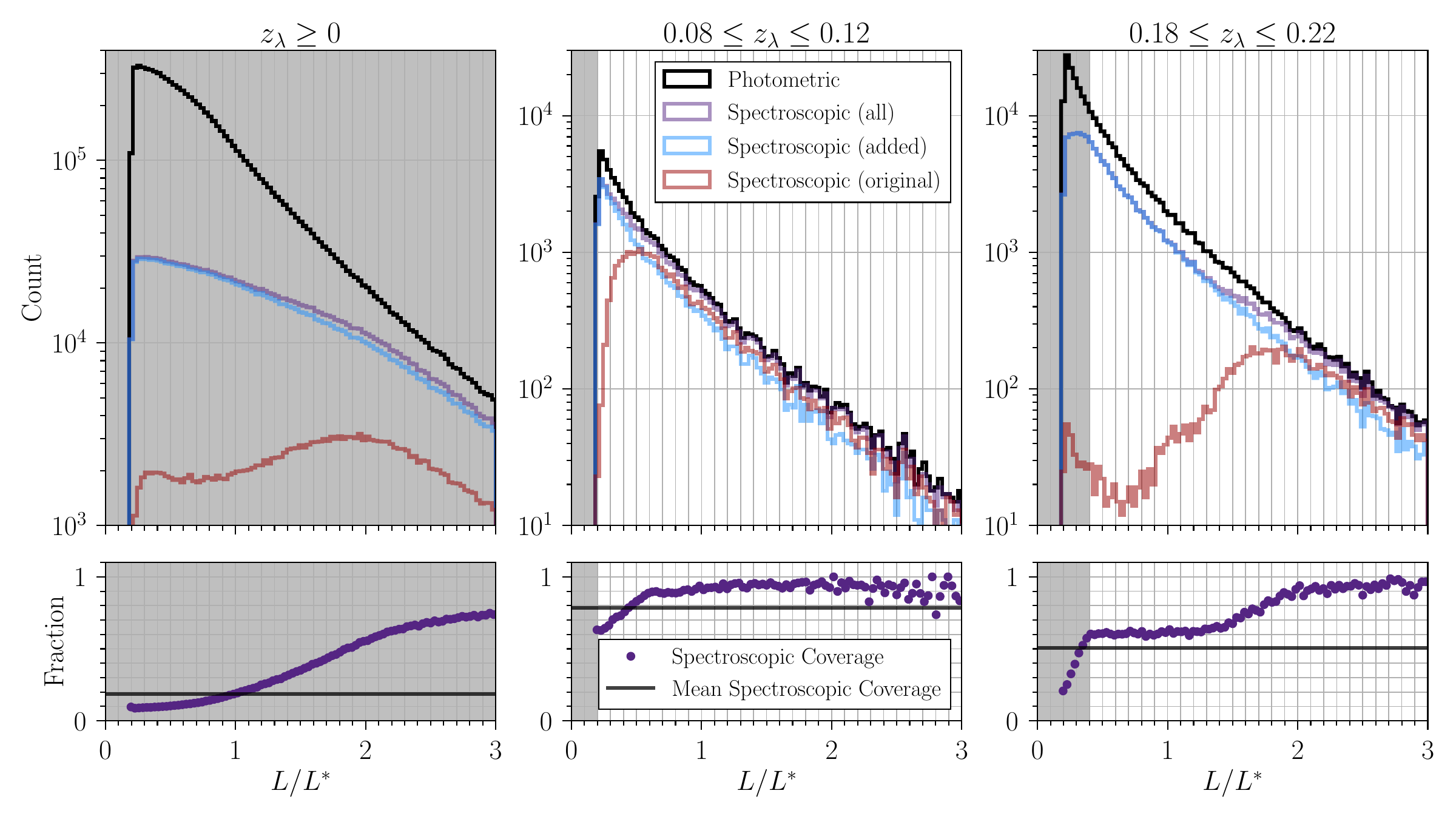}
    \caption[Luminosity distribution of cluster galaxy samples]{Luminosity distribution of \redmapper{} cluster galaxies, inferred assuming \redmapper{} photometric redshifts and $i$-band magnitudes. The upper left, middle, and right panels illustrate \redmapper{} member galaxies for clusters with photometric redshifts satisfying $\zlambda>0$, $0.08 \leq \zlambda \leq0.12$, and $0.18 \leq \zlambda \leq 0.22$, respectively. The red curve corresponds to the SDSS DR9 spectroscopy used to construct the \redmapper{} catalogue; the blue curve corresponds to spectra used in this work that were not used by  \redmapper{} (i.e., DESI DR2 and SDSS DR17 exclusive of data in DR9); the purple combines the spectroscopic data represented by the blue and red curves. Lower panels indicate the fraction of galaxies with spectroscopic redshifts, where the horizontal line indicates the sample mean. For the analysis in this work, we adopt lower luminosity limits of $0.2\lstar$ at $z\sim 0.1$ and $0.4\lstar$ at $z\sim0.2$; for details on this choice see \S \ref{sec:data}. The grey shading indicates regions of selection-parameter space not used in this work. This figure illustrates that new spectroscopic observations improve cluster galaxy coverage, enabling the measurements in this work.
    }
    \label{fig:sdss_desi_coverage}
\end{figure*}

This paper is organized as follows. In \S \ref{sec:data} we describe the spectroscopic data used to characterize \redmapper{} galaxy clusters. In \S \ref{sec:modeling} we describe our model for galaxy cluster line-of-sight velocity. We present our measurements of stacked velocity dispersion and projection effects in \S \ref{sec:results} and robustness tests thereof in \S \ref{sec:robustness_tests}. We present conclusions from our measurements and make recommendations on the use of \redmapper{} in \S \ref{sec:conclusions}. Finally we discuss our results in the context of other work and we highlight future directions of research in \S \ref{sec:discussion}. 
A flat \lcdm{} cosmology with $H_0 = 70 \mathrm{km\  s}^{-1} \mathrm{Mpc}^{-1}$ and $\Omega_{\rm{m}}=0.3$ is assumed throughout. 
\section{Data}
\label{sec:data}

We use galaxy clusters identified from the Sloan Digital Sky Survey Data Release 8 (SDSS DR8) imaging consisting of 14,000 deg$^2$ observed with the 2.5-m telescope at Apache Point Observatory \citep{sdss_dr8}. After quality criteria are applied to the data, 10,500 deg$^2$ of imaging remain. The galaxy cluster identification algorithm additionally uses the SDSS DR9 spectroscopic catalogue containing 1.3 million galaxy spectroscopic redshifts \citep{sdss_spec}. The primary galaxy cluster catalogue used in this analysis was produced by running \redmapper{} with these imaging and spectroscopic datasets as inputs as described in \citet{redmapper1, redmapper4}. This catalogue provides an optimal combination of area and completeness at the relevant redshift to enable the systematic study of algorithmic performance via representative spectroscopy. 

The Dark Energy Spectroscopic Instrument survey provides a three-dimensional map of the Northern Hemispheric sky constructed from spectroscopic galaxy observations collected with the dedicated 4-m Mayall Telescope at Kitt Peak National Observatory. DESI represents a significant improvement over SDSS due in large part to its robotic $\sim$5000 fibre positioning system. 

In this study we supplement the galaxy spectroscopy used to construct the SDSS DR8 \redmapper{} catalogue with spectra collected subsequently by SDSS (Data Release 17) and DESI (Data Release 2). While cluster galaxies are not a target population per se of the main DESI surveys, many DESI targets are cluster members.  The result yields unprecedented spectroscopic coverage of galaxy cluster members for tens of thousands of galaxy clusters and groups. While DESI coverage for individual clusters is limited by fibre collision ($\sim2\farcm{4}$, \citealt{Martini2018, FocalPlane.Silber.2023}), stacking many clusters in bins of optical richness enables characterizing the mean performance of the algorithm as a function of relevant properties such as richness and redshift. 

We apply the following selection criteria to these data to produce an analysis sample with sufficiently low catastrophic redshift failures, following similar cuts made by \citet{Ross2025}.

\begin{center}
\begin{varwidth}{\textwidth}
\begin{enumerate}[label=\roman*)]
\item \texttt{SPECTYPE} $=$ \texttt{GALAXY}
\item \texttt{ZWARN} $= 0$
\item \texttt{DELTACHI2} $> 40$.
\end{enumerate}
\end{varwidth}
\end{center}

Here \texttt{SPECTYPE} indicates the spectral type of the template best fit to the DESI observed spectrum, \texttt{ZWARN} is a quality flag indicating a known problem with a given spectroscopic redshift fit \citep[for details see][]{SurveyOps.Schlafly.2023, Spectro.Pipeline.Guy.2023}, and \texttt{DELTACHI2} is a difference between two $\chi^2$ values: the $\chi^2$ metric between the best-fitting spectral template with respect to the data and the $\chi^2$ metric between the \textit{second} best-fitting spectral template with respect to the data. 
\begin{figure}
    \centering
    \includegraphics[width=0.5\textwidth]{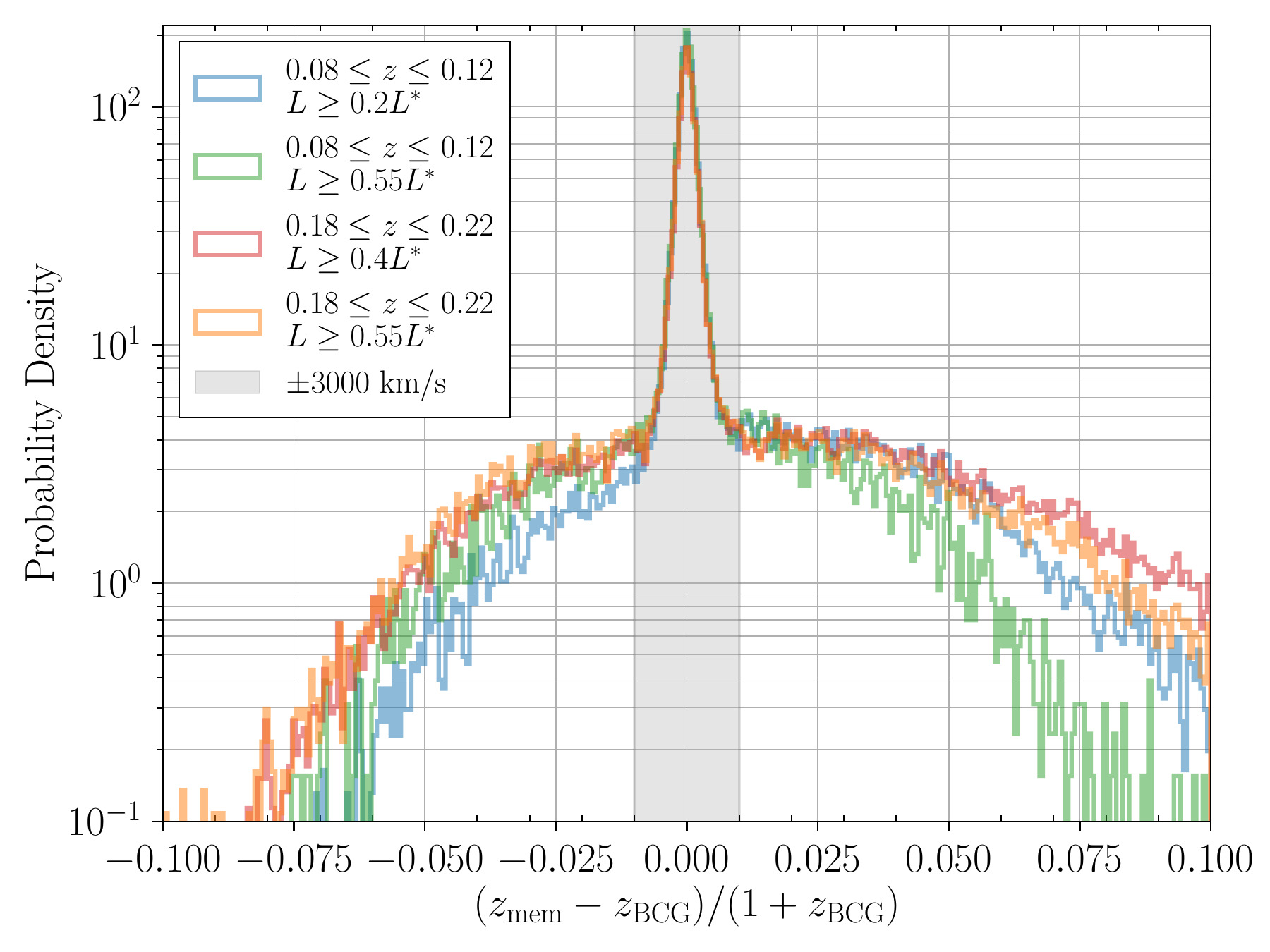}
    \caption[Cluster galaxy line-of-sight velocities]{Line-of-sight velocity distribution of \redmapper{} cluster galaxies for the samples used in this analysis. This distribution exhibits evidence of multiple contributing components, representing true cluster halo members and galaxies in projection, respectively. The galaxies in projection can in principle be attributed to multiple failure modes dominated by photometric redshift uncertainty, and including blending, etc. Modeling this distribution enables both calibration of \redmapper{} richness and direct use of stacked cluster velocity dispersion for cosmological studies.}
    \label{fig:v_los_dens_data}
\end{figure}

\begin{table*}
    \begin{tabular}{llllrrrrrr}
    \toprule
    Type & Cluster &  Galaxy & & \multicolumn{6}{|c|}{Richness} \\
    & Redshift & Luminosity &  &  5--20 & 20--27.9 & 27.9--37.6 & 37.6--50.3 & 50.3--69.3 & 69.3--140 \\    
    \midrule\midrule
 Fiducial & \multirow[t]{2}{*}{ 0.08-0.12 } & \multirow[t]{2}{*}{ $L / L^{*} \geq 0.20$ }  & Clusters & 3,390 & 183 & 84 & 51 & 26 & 12 \\
        & &        & Galaxies & 34,679 & 4,541 & 2,815 & 2,214 & 1,430 & 882 \\
\midrule
 Conservative & \multirow[t]{2}{*}{ 0.08-0.12 } & \multirow[t]{2}{*}{ $L / L^{*} \geq 0.55$ } & Clusters & 3,317 & 183 & 84 & 51 & 26 & 12 \\
        & &       & Galaxies & 15,546 & 2,054 & 1,316 & 1,030 & 695 & 445 \\
\midrule
\midrule
 Fiducial & \multirow[t]{2}{*}{ 0.18-0.22 } & \multirow[t]{2}{*}{ $L / L^{*} \geq 0.40$ } & Clusters & 10,931 & 609 & 293 & 179 & 82 & 32 \\
        & &       & Galaxies & 61,121 & 7,195 & 4,521 & 3,395 & 2,157 & 1,107 \\
\midrule
 Conservative & \multirow[t]{2}{*}{ 0.18-0.22 } & \multirow[t]{2}{*}{ $L / L^{*} \geq 0.55$ } & Clusters & 10,669 & 609 & 293 & 179 & 82 & 32 \\
        & &       & Galaxies & 42,637 & 5,105 & 3,257 & 2,424 & 1,555 & 825 \\

    \bottomrule

    \end{tabular}
    \caption[Cluster Galaxy Samples]{Main cluster galaxy samples analyzed in this work. For each galaxy in this analysis, we have a spectroscopic redshift from either SDSS DR17 or DESI DR2. We spectroscopically characterize \redmapper{} cluster galaxies at two redshifts $z\sim0.1$ and $z\sim0.2$, down to multiple luminosity thresholds in order to test the dependence of projection effects as a function of redshift and galaxy brightness. DESI substantially increases the available data for this analysis, enabling a first empirical test of projection effects with these redshift and luminosity limits. At redshifts $z\sim0.1$ and $z\sim0.2$ the new spectroscopy increases the number of cluster galaxies with redshifts by factors of $\sim1$ and $\sim10$, respectively.} \label{tab:sample}
    \end{table*}

We match the SDSS DR8 \redmapper{} galaxy cluster member catalogue with both the DESI DR2 redshift catalogue and the SDSS DR17 spectroscopic dataset using a matching radius of $1\farcs{5}$ around each putative \redmapper{} cluster member galaxy. The fractional spectroscopic coverage of \redmapper{} cluster galaxies before and after including the SDSS DR17 and DESI DR2 data is shown in Figure \ref{fig:sdss_desi_coverage}. Luminosity $L/\lstar$ for each \redmapper{} cluster galaxy, as shown in this Figure, is determined uniformly for the entire sample with the \redmapper{} cluster photo-$z$ ($z_{\lambda}$) and the galaxy $i$-band magnitude following the prescription of \citet{redmapper1}.

At redshift $z\sim0.1$, the SDSS main galaxy sample alone is complete for galaxies brighter than $r\leq 17.77$ ($L\geq0.55\lstar$), but has rapidly diminishing coverage completeness for fainter galaxies \citep{strauss2002}. DESI provides data down to the faintest limit included in the \redmapper{} catalogue ($L\geq 0.2\lstar$). At redshift $z\sim0.2$, the SDSS data used for constructing the \redmapper{} catalogue have relatively few galaxy spectra. DESI coverage of \redmapper{} cluster galaxies is fairly uniform above $L\geq0.4\lstar$ at this redshift at $\gtrsim 60$ per cent. Aside from the luminosity-dependent transition between the bright regime where SDSS contributes substantially and the faint regime where DESI dominates ($L\sim 1.5 \lstar$), there is no strong luminosity dependence in the spectroscopic coverage above $L>0.4\lstar$. This is exemplified by the small deviations in coverage fraction between luminosity bins in the range $0.4 \leq L/\lstar \leq 1$ ($\sigma_f = 0.009$, i.e. 1.5 per cent of the mean value $\langle f \rangle = 0.60$). These coverage statistics enable a study of projection effects at $z\sim0.1$ down to the faintest \redmapper{} galaxies ($L\geq 0.2\lstar$) as well as at $z\sim0.2$ down to moderate luminosities $L\geq 0.4\lstar$. We adopt these as our fiducial lower luminosity limits at $z\sim0.1$ and $z\sim0.2$, respectively. We select clusters with \redmapper{} photometric redshifts $\zlambda$ satisfying $0.08 \leq \zlambda \leq 0.12$ and $0.18 \leq \zlambda \leq 0.22$ to probe these two redshift regimes, respectively. The selected cluster samples are divided into the six richness bins of \citetalias{Myles2021} for consistency with past measurements. In addition to these fiducial lower limits we perform our analysis for conservative samples satisfying $L\geq0.55\lstar$ for baseline comparisons. 

The line-of-sight velocities for these samples are shown in Figure \ref{fig:v_los_dens_data}. We truncate the data satisfying $|\frac{\Delta z}{1+z}|>0.1$. This cut removes 43 and 613 (1049 and 2947) galaxies with extreme velocities ($v > 30,000 \ \mathrm{km \  s}^{-1}$) at $z\sim0.1$ ($z\sim0.2$) for the conservative and fiducial samples, respectively.

The sizes of our samples after matching the \redmapper{} and spectroscopic catalogues are shown in Table \ref{tab:sample}. The number of selected clusters is set by how many \redmapper{} cluster detections have both a central galaxy spectroscopic redshift and at least one valid member spectroscopic redshift. For the fiducial luminosity cuts, these are a subset of the 4,389 
and 
15,305
candidate \redmapper{} clusters (and groups) at redshifts in our $z\sim0.1$ and $z\sim0.2$ bins, respectively.  In other words, roughly 85 and 80 per cent of the \redmapper{} cluster candidates at $z\sim0.1$ and $z\sim0.2$ are included in our analysis. While for each individual cluster DESI coverage is not complete, combining clusters in bins of richness and redshift provides well sampled cluster galaxy distributions (i.e. distributions comprising hundreds to tens of thousands of galaxies). 

\section{Modeling}
\label{sec:modeling}

\subsection{Modeling galaxy line-of-sight velocity}
The stacked line-of-sight velocity distribution of \redmapper{} clusters exhibits multiple components, as shown in Figure \ref{fig:v_los_dens_data}. Noting that the Maxwell-Boltzmann distribution for the velocity vector is a multivariate Gaussian with mean zero, we expect the true cluster halo galaxies to be normally distributed in $\frac{\Delta z}{1+z}$, consistent with the observed distribution of Figure \ref{fig:v_los_dens_data} within 3,000 km s$^{-1}$. 
The galaxy population in projection is dominated by those in correlated structures along the line-of-sight whose photometric redshifts make them indistinguishable from true cluster halo galaxies. \citetalias{Myles2021} found this distribution to also be fit with a Gaussian with mean consistent with zero when all data were combined. Visual inspection of the samples in Figure \ref{fig:v_los_dens_data} shows that as fainter galaxies are included in the sample, the distribution of galaxies in projection is driven to higher mean redshift than that of the clusters, with possible evidence of asymmetrically greater incidence of projections at higher redshifts. This is unsurprising, as it is consistent with projections being more likely to come from fainter galaxies than brighter galaxies, all else being fixed. We identify the cause of the asymmetry as being due fundamentally to a combination of photometric noise and differential volume. For a given cluster, galaxies of a fixed luminosity are fainter when on the far side rather than the near side relative to the observer. Redshift differentials of fixed value correspond to spherical shells of larger volume as redshift increases. We discuss this further in \S \ref{sec:results_lum_dep}. Upon further investigation, the extreme ends of the non-Gaussian wings can be attributed to multiple uncommon failure modes including blended galaxy detections with colours biased ipso facto and galaxies with very low \redmapper{} cluster membership probabilities.

Recognizing that our primary goal is characterization of the clusters, we impose a $| \frac{\Delta z}{1+z}| < 0.1 = 30,000$ km s$^{-1}$ cut on the data, in excess of reasonable physical priors on cluster galaxy velocities  by at least a full order of magnitude ($<$3,000 km s$^{-1}$). This cut removes extreme outliers primarily responsible for the deviation of the overall distribution from Gaussianity, enabling continued use of the simple double-Gaussian model for the data. 

\begin{figure*}
\includegraphics[width=0.75\textwidth]{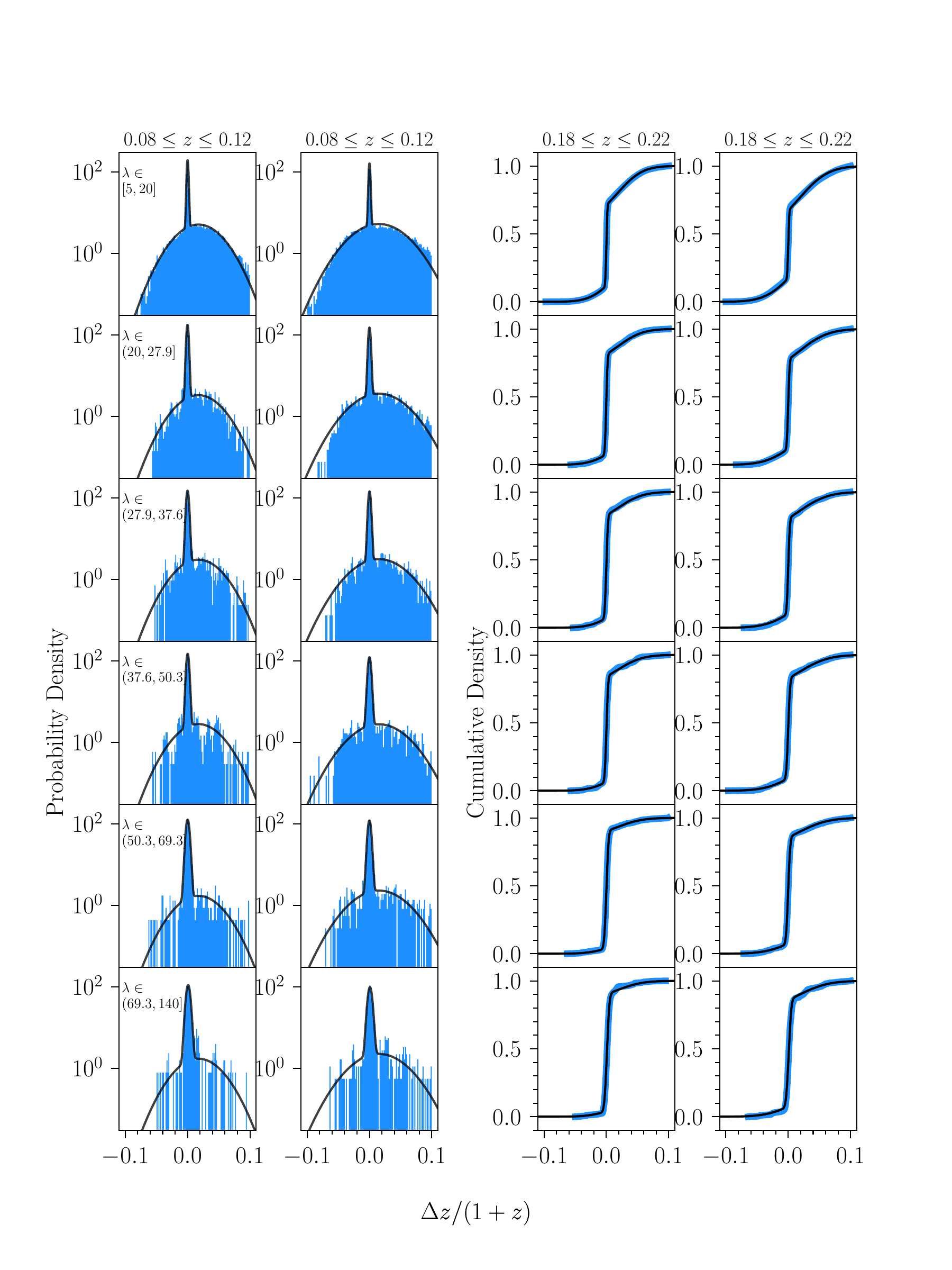}
    \caption[Model fit to data]{Line-of-sight velocity distribution of \redmapper{} galaxy clusters for the fiducial samples at $z
\sim0.1$ and $z\sim0.2$, respectively. A double Gaussian model is fit to the data, where the component representing galaxies in projection is constrained jointly with all richness bins. The galaxies in projection exhibit mean redshift on the far side of their respective host clusters and asymmetry in their distribution, with a surplus (shortage) of data observed relative to the model at the extreme end on the far (near) side of the cluster, discussed further in \S \ref{sec:results}. Outlying galaxies with $|\frac{\Delta z}{1+z}|>0.1$ are excluded from the samples used in this analysis. This model is used to quantify the degree of richness- and redshift- dependence in the projected galaxy fraction parameter $\fproj$, as shown in Figure \ref{fig:fproj_vs_lambda}.}\label{fig:vlos_dens_data_and_model_richness_bins}
\end{figure*}

Our model for the putative cluster galaxy line-of-sight velocities is:

\begin{eqnarray} \label{eqn:halo_model}
    p(\deltaznormed) &=& f_{\mathrm{cl}} \ \mathcal{N}(\deltaznormed|\ \mu_{\mathrm{cl}}, \sigma_\mathrm{cl}) \\
    && + \, f_{\mathrm{proj}} \ \mathcal{N}(\deltaznormed|\ \mu_{\mathrm{proj}}, \sigma_{\mathrm{proj}}). \nonumber
\end{eqnarray}

Here $f_{\mathrm{cl}}$ represents the fraction of putative galaxies which are true cluster members and $f_{\mathrm{proj}} \equiv 1 - f_{\mathrm{cl}}$ represents the corresponding amplitude of projection effects, after truncation of extreme projections. 

\subsection{Modeling richness bias due to projection effects}
\label{sec:lambdabias}

Because \redmapper is designed to identify overdensities of \textit{red-sequence} galaxies from photometric data, modeling richness bias must take galaxy colour into account. We use the $\lambdatrue$ richness estimate for \redmapper{} clusters developed in \citetalias{Myles2021}. 

This richness definition is defined to achieve two goals: first, $\lambdatrue$ should be similar enough to $\lambdaobs$ that a comparison of the two measures the extent to which \redmapper is affected by projection effects; second, to be most useful for subsequent cosmological analyses, $\lambdatrue$ should relate to the cluster mass as simply as possible, with minimal intrinsic scatter. 

\subsubsection{\redmapper richness}
\redmapper defines a probability ($\pmem$) that a galaxy is a red cluster member above a threshold in $L/\lstar$. The \redmapper richness of a given cluster is defined as the sum of $\pmem$ over cluster members:

\begin{equation}
\label{eqn:lambdaobs}
\begin{split}
\lambdaobs &\equiv  \sum_{\text{mem}} p(\text{galaxy is a red cluster member} | \text{photometry}) \\
&= \sum_{\text{mem}} p(\vec{x}|\lambda) \ p_{\text{free}} \ \theta_r \ \theta_i.
\end{split}
\end{equation}

This probability that a galaxy is a red member of a given cluster is defined as a product of four factors \citep{redmapper4}:

\begin{enumerate}
    \item $p(\vec{x}|\lambda)$: the probability that a galaxy with observed photometric properties $\vec{x}=(g-r, r-i, i-z, i, \alpha, \delta)$ (multiple photometric colours, \textit{i}-band magnitude, and position on the sky in right ascension and declination) is a red member of a cluster at that position of richness $\lambda$. This term is evaluated with a matched filter that comprises three sub-filters: the cluster galaxy radial number density profile, the cluster luminosity function, and a $\chi^2$ measure of the consistency of the galaxy colour with the red-sequence model at a given redshift;
    \item $p_{\text{free}}$ is the probability that the galaxy does not belong to a previous cluster in the percolation step of the algorithm;
    \item $\theta_r$ is a radial weight function that acts as a smooth radial threshold to account for the small astrometric uncertainty on the position of a given candidate member;
    \item $\theta_i$ is an \textit{i}-band magnitude weight function that acts as a smooth luminosity threshold at $0.2 \lstar$ (under the assumption of a photometric redshift) to account for the photometric uncertainty on the apparent $i$-band magnitude of a given candidate member.
\end{enumerate}

\subsubsection{Spectroscopic Richness}
We employ an improved richness estimate, $\lambdatrue$, defined by \citetalias{Myles2021} as:

\begin{equation}
\label{eqn:lambdaspec}
\begin{split}
\lambdatrue &\equiv  \sum_{\text{mem}} p(\text{galaxy is a red member} | \text{spectroscopy, photometry}) \\
&= \sum_{\text{mem}} \pspec \pred p_{\mathrm{free}} \theta_r \theta_i.\\
\end{split}
\end{equation}

In this expression, $\pspec$ is given by,

\begin{equation}
\label{eqn:pspec}
\pspec = \frac{f_{\mathrm{cl}} \ \mathcal{N}(\deltaznormed|\ \mu_{\mathrm{cl}}, \sigma_\mathrm{cl})}{f_{\mathrm{cl}} \ \mathcal{N}(\deltaznormed|\ \mu_{\mathrm{cl}}, \sigma_\mathrm{cl}) + \, f_{\mathrm{proj}} \ \mathcal{N}(\deltaznormed|\ \mu_{\mathrm{proj}}, \sigma_{\mathrm{proj}})},
\end{equation}

and $\pred$ is the probability of a galaxy being drawn from the \redmapper{} red-sequence model population, given its photometry. Following \citealt{redmapper4}, 

\begin{equation}
    \pred (\chi_{\mathrm{s}}) = \frac{1}{2} \left[ 1 - \mathrm{erf} \left( \frac{\ln(\chi_{\mathrm{s}}/\chi_{\mathrm{ref}})}{\sqrt{2}\sigma} \right) \right].
\end{equation}

This definition of $\lambdatrue$ makes use of both spectroscopic and photometric information to determine membership probability; it additionally accounts for galaxy position and cluster percolation in the same way as \redmapper, thereby facilitating comparison with \redmapper richness.
Combining $\pspec$ with $\pred$ does differ from the \redmapper-defined $\pmem$ because the matched filter that \redmapper uses to determine $p(\vec{x}|\lambda)$ contains sub-filters for the cluster density profile and the cluster luminosity function. 

We estimate the richness bias due to projection effects by comparing $\lambdaobs$ with $\lambdatrue$ in each richness bin. For each richness bin $j$, the richness bias from the selected candidate members $m$ is given by:
\begin{align}
\label{eqn:projeffects}
    b_{\lambda,j} &\equiv \dfrac{\sum\limits_{m \in j}{\pmem} - \sum\limits_{m \in j}{\pspec \pred p_{\mathrm{free}} \theta_r \theta_i}}{\sum\limits_{m \in j}{\pmem}} \nonumber \\ 
    &= \dfrac{\sum \lambdaobs^{L\geq L_{\text{cut}} } - \sum \lambdatrue^{L\geq L_{\text{cut}}}}{\sum \lambdaobs^{L\geq L_{\text{cut}}}}. 
\end{align}

\begin{figure*}
    \centering
    \includegraphics[width=0.95\textwidth]{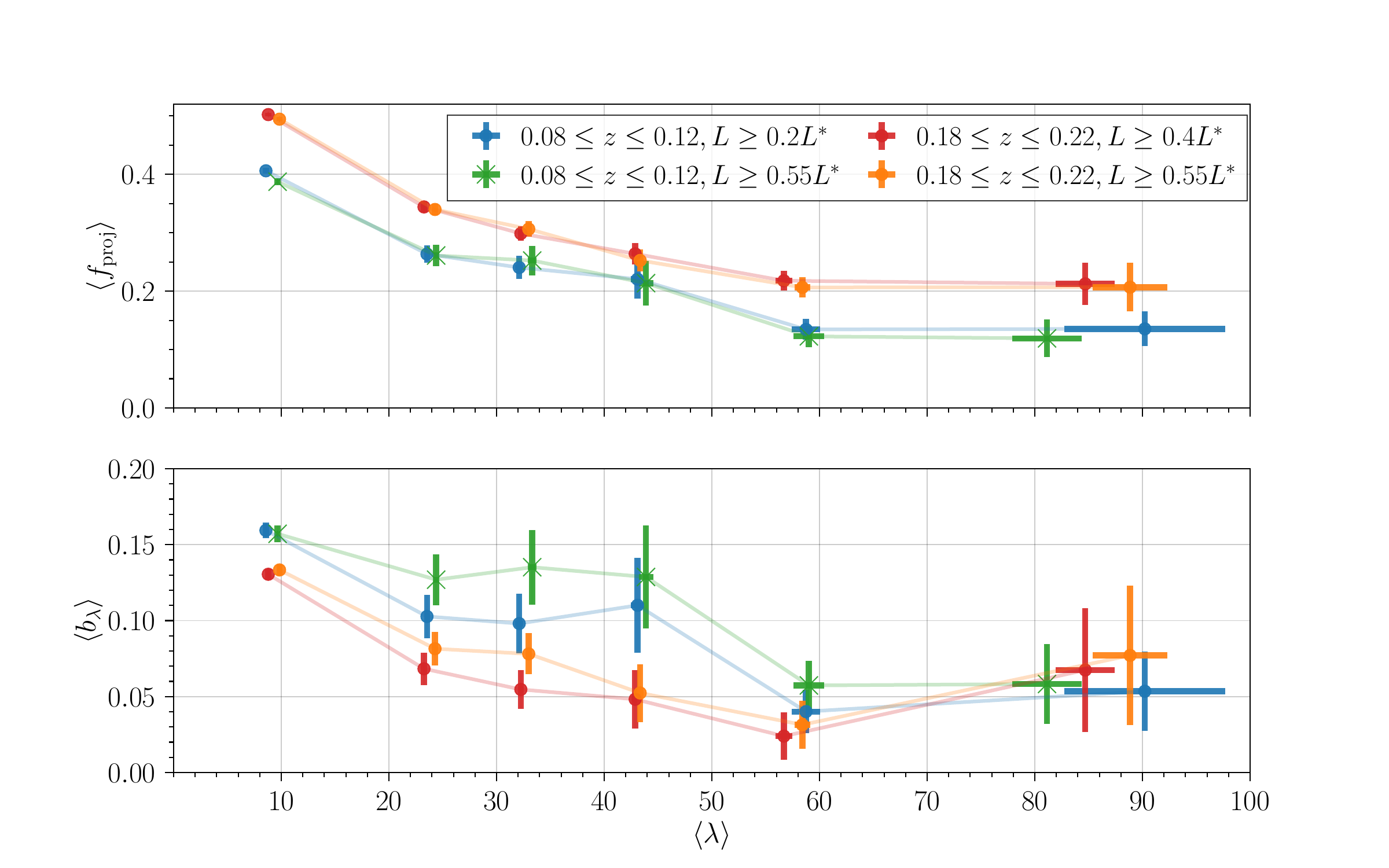}
    \caption[Richness and redshift dependence of projection effects]{Dependence of projection effects on \redmapper{} richness ($x$-axis) and redshift. Shown in the top panel is the value of the model parameter $\fproj$ encoding the amplitude of projection effects. $\langle \fproj \rangle$ is the mean fraction of putative \redmapper{} galaxies whose line-of-sight velocities indicate that these galaxies are not true cluster halo members. The bottom panel illustrates the richness bias $b_\lambda$ derived from the probability of each individual cluster galaxy being in projection according to our model. The shape and relative amplitudes of the Gaussian mixture model components attributed to projection effects can vary for different samples; $\fproj$ summarizes the total amplitude of projection effects. This figure illustrates the redshift and richness dependence of projection effects. Small $x$-axis offsets have been introduced to reduce overlap between points.
    }\label{fig:fproj_vs_lambda}
\end{figure*}

\subsection{Model fitting procedure}
\label{sec:sampling}
Using Equation \ref{eqn:halo_model}
as a Bayesian likelihood function and broad uniform priors on the model parameters, we first evaluate the posterior of the model parameters given each data set of line-of-sight velocities by sampling the posterior probability function of the data with the Metropolis-Hastings Markov Chain Monte Carlo (MCMC) algorithm as implemented in \texttt{emcee}  \citep{emcee}. To account for uncertainty in model parameter constraints (including uncertainty due to sample variance), we implement bootstrap re-sampling of the selected clusters in each cluster richness bin. We maximize the posterior for each bootstrap sample via Nelder-Mead minimization of the negative log-likelihood with initialization of the optimization starting at values from the Markov chain. For this fiducial model fit, the mean parameter for the cluster galaxy distribution is fixed to zero (for details on this choice, see Appendix \ref{sec:model_validation}). 
The final model best-fitting parameters and corresponding uncertainties reported are computed from the bootstrap samples, with 30,000 samples for the fiducial fits. 

\section{Results}
\label{sec:results}

We fit the model specified in \S \ref{sec:modeling} to the data sample selections defined in \S \ref{sec:data}. The results of this procedure for the fiducial samples are shown in Figure \ref{fig:vlos_dens_data_and_model_richness_bins} and summarized in Table \ref{tab:parameters}. 

Overall, we find the model accounts for the observed data for the majority of parameter space in line-of-sight velocity. The excess of data with respect to the model on the far side of the cluster is driven by the fainter galaxies in the sample, as illustrated in Figure \ref{fig:v_los_dens_data_lum}. Our primary goal is the characterization of the Gaussian representing cluster galaxies, as well as the relative amplitudes of this model component and its counterpart of projected galaxies. Since the relative amplitudes of these components are relatively insensitive to the relatively few galaxies at the extreme ends of line-of-sight velocity, we proceed with this model for our analysis. 

Repeating the model fitting for samples with varying redshift, lower luminosity, and richness enables a determination of the amplitude of projection effects as a function of these variables. The fraction of putative \redmapper{} galaxies whose spectroscopic information indicates they are likely in projection ranges from $\sim10$ to $\gtrsim 50$ per cent, with significant variation as a function of richness, redshift, and luminosity. 

In addition to the projection fraction $\fproj$ we constrain the richness bias $b_\lambda$. The latter represents the degree to which, on average, \redmapper{} richness is boosted for a given bin due to galaxies in projection; this quantity is smaller than $\fproj$ for any given bin because it accounts for the membership probability ($0 \leq \pmem \leq 1$) of each cluster galaxy. 

\subsection{Richness dependence of projection effects}
\label{sec:results_richness_dep}

As shown in Figure \ref{fig:fproj_vs_lambda}, there is a strong dependence of the amplitude of projection effects with richness, introducing a corresponding bias in the mass-richness relation. We reinforce the findings of \citetalias{Myles2021} by measuring projection effects to fainter luminosities and at higher redshifts. 
Using the additional DESI data to test the degree of richness dependence at higher redshift ($z\sim0.2$) than before, we find that the richness dependence of projection effects previously observed at $z\sim0.1$ persists at $z\sim0.2$.
In \S \ref{sec:discussion} we discuss the implications of this richness dependence for cosmology in light of the fact that the median cluster catalogue redshift (from imaging data, irrespective of spectroscopic coverage) is $z\sim0.4$ ($z\sim0.5$) for the $5 \leq \lambda \leq 140$ ($20 \leq \lambda \leq 140$) sample. 

We also note that the $z\sim0.1, L\geq 0.55 \lstar$ sample used in \citetalias{Myles2021} demonstrated a non-monotonic trend in the amplitude of projection effects ($\fproj$) as a function of richness, where the best-fitting value for the $27.9 < \lambda \leq 37.6$ richness bin was slightly higher than that of the second richness bin (although not statistically significantly so). With the increased data sample of this work, the trend of projection effects with richness is monotonically decreasing. We therefore find that the deviation from monotonicity found in \citetalias{Myles2021} is consistent with uncertainty due to sample variance. 

\subsection{Redshift dependence of projection effects}
\label{sec:results_redshift_dep}

\begin{figure}
    \centering
    \includegraphics[width=0.49\textwidth]{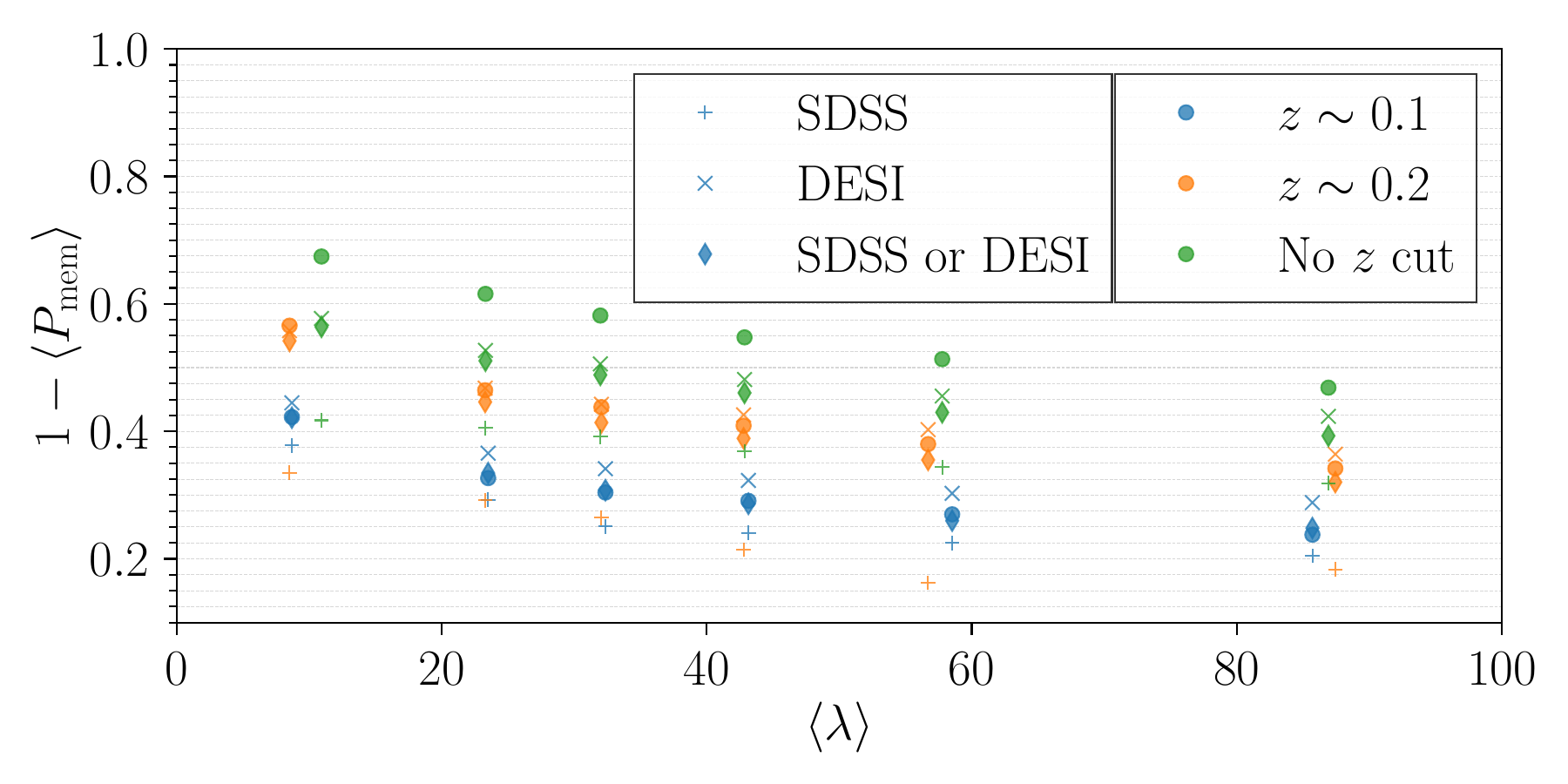}
    \caption[\redmapper{} membership probability dependence on brightness]{Mean \redmapper{} membership probability as a function of richness. Circular markers indicate all members, whereas non-circular markers indicate the subsets with spectroscopic data, as specified in the left legend. On average, $\pmem$ decreases as redshift increases from $z\sim0.1$ to $z\sim0.2$. This effect is observed in the photometric sample and in the subsamples with representative spectroscopy from SDSS and DESI. The decrease in mean $\pmem$ as redshift increases counteracts the increase in the projection fraction of \redmapper{} galaxies. Taking both of these effects into account, richness bias improves from $z\sim0.1$ to $z\sim0.2$. Determining how richness bias scales fully with redshift, however, depends on spectroscopic measurement of $\fproj$ at all redshifts.}\label{fig:mean_pmem}
\end{figure}

The high coverage of DESI data enables a first study of the redshift dependence of optical cluster projection effects over a large footprint. Comparing the $z\sim0.1$ and $z\sim0.2$ samples shown in Figure \ref{fig:fproj_vs_lambda}, a clear difference is apparent. The higher redshift clusters have a larger fraction of putative galaxies that are projection effects. Moreover, the richness dependence of $\fproj$ is similar to that observed at lower redshift. These results are not surprising as they are consistent with the simple explanation that projection effects are driven by photometric redshift uncertainty, which generally increases for the higher redshift sample compared to the lower redshift sample at a given intrinsic luminosity. However, the fact that projection effects can become so much more severe ($\sim25$ per cent larger in the worst case) for such a small increase in redshift is notable. Very little cosmological evolution in the cluster galaxy population is expected over this redshift increment,  heightening the importance of empirical constraints on projection effects.

The richness bias $b_\lambda$ exhibits a different trend with richness. Namely, at high richness there is little to no evidence for redshift dependence in richness bias and at low richness the richness bias appears lower at $z\sim0.2$ than at $z\sim0.1$. The clear redshift dependence in $\fproj$ with richness contrasted with the reduced trend in $b_\lambda$ can be attributed to, in general, \textit{lower} average \redmapper{} membership probabilities at $z\sim0.2$ than at $z\sim0.1$. As shown in Figure \ref{fig:mean_pmem}, higher redshift samples have lower average membership probabilities (illustrated by points higher on the $y$-axis). This counteracts the larger fraction of projected galaxies. The mean $\pmem$ values at $z\sim0.1$ are approximately 10 to 26 per cent larger than for $z\sim0.2$, which is comparable and in some bins in excess to the fractional increase in projected fraction from the lower to the higher redshift bin. Notwithstanding the known average \redmapper{} probabilities as a function of photometric redshift, it is not guaranteed that the improvement of richness bias from $z\sim0.1$ to $z\sim0.2$ would persist or continue at higher redshifts, since richness bias also depends on the projection fraction. 

\begin{table*}
\begin{center}
\begin{tabular}[h]{c l |c c c c c c }


Sample & Parameter & $5\leq \lambda \leq$ 20 & $20<\lambda\leq27.9$ &  $27.9<\lambda\leq37.6$ & $37.6<\lambda\leq50.3$ & $50.3<\lambda\leq69.3$ & $69.3<\lambda\leq140$ \\
\hline
\hline
$z\sim0.1$ & $\langle f_{ \text{proj.} } \rangle $ & 0.406 $\pm$ 0.005 & 0.263 $\pm$ 0.015 & 0.240 $\pm$ 0.020 & 0.220 $\pm$ 0.033 & 0.134 $\pm$ 0.018 & 0.135 $\pm$ 0.030\\
\multirow[t]{2}{*}{ $ L\geq 0.2 L^{*}$ } & $\langle \sigma_{ \text{cl.} } \rangle$ [ km s$^{-1}$ ]  & 373 $\pm$ 5 & 509 $\pm$ 14 & 615 $\pm$ 24 & 638 $\pm$ 27 & 822 $\pm$ 42 & 962 $\pm$ 111\\
 & $\langle b_{ \lambda } \rangle$  & 0.159 $\pm$ 0.005 & 0.103 $\pm$ 0.014 & 0.098 $\pm$ 0.020 & 0.110 $\pm$ 0.031 & 0.040 $\pm$ 0.014 & 0.054 $\pm$ 0.026\\
\hline$z\sim0.1$ & $\langle f_{ \text{proj.} } \rangle$ & 0.387 $\pm$ 0.006 & 0.261 $\pm$ 0.018 & 0.252 $\pm$ 0.025 & 0.214 $\pm$ 0.038 & 0.123 $\pm$ 0.018 & 0.119 $\pm$ 0.032\\
\multirow[t]{2}{*}{ $ L\geq 0.55 L^{*}$ } & $\langle \sigma_{ \text{cl.} } \rangle$ [ km s$^{-1}$ ]  & 375 $\pm$ 6 & 507 $\pm$ 17 & 607 $\pm$ 26 & 637 $\pm$ 29 & 772 $\pm$ 38 & 948 $\pm$ 78\\
  & $\langle b_{ \lambda } \rangle$  & 0.157 $\pm$ 0.005 & 0.127 $\pm$ 0.017 & 0.135 $\pm$ 0.025 & 0.129 $\pm$ 0.034 & 0.057 $\pm$ 0.016 & 0.058 $\pm$ 0.026\\
\hline$z\sim0.2$ & $\langle f_{ \text{proj.} } \rangle$ & 0.502 $\pm$ 0.003 & 0.344 $\pm$ 0.010 & 0.299 $\pm$ 0.012 & 0.264 $\pm$ 0.018 & 0.218 $\pm$ 0.017 & 0.212 $\pm$ 0.036\\
\multirow[t]{2}{*}{ $ L\geq 0.4 L^{*}$ }  & $\langle \sigma_{ \text{cl.} } \rangle$ [ km s$^{-1}$ ]  & 380 $\pm$ 3 & 530 $\pm$ 11 & 596 $\pm$ 17 & 728 $\pm$ 29 & 788 $\pm$ 33 & 988 $\pm$ 67\\
 & $\langle b_{ \lambda } \rangle$  & 0.131 $\pm$ 0.003 & 0.068 $\pm$ 0.011 & 0.055 $\pm$ 0.013 & 0.048 $\pm$ 0.019 & 0.024 $\pm$ 0.016 & 0.067 $\pm$ 0.041\\
\hline$z\sim0.2$ & $\langle f_{ \text{proj.} } \rangle$ & 0.494 $\pm$ 0.003 & 0.340 $\pm$ 0.010 & 0.306 $\pm$ 0.014 & 0.252 $\pm$ 0.019 & 0.206 $\pm$ 0.017 & 0.207 $\pm$ 0.042\\
\multirow[t]{2}{*}{ $ L\geq 0.55 L^{*}$ }  & $\langle \sigma_{ \text{cl.} } \rangle$ [ km s$^{-1}$ ]  & 383 $\pm$ 4 & 534 $\pm$ 13 & 597 $\pm$ 19 & 751 $\pm$ 31 & 801 $\pm$ 36 & 974 $\pm$ 70\\
 & $\langle b_{ \lambda } \rangle$  & 0.133 $\pm$ 0.003 & 0.081 $\pm$ 0.011 & 0.078 $\pm$ 0.014 & 0.052 $\pm$ 0.019 & 0.031 $\pm$ 0.016 & 0.077 $\pm$ 0.046\\
\hline\rule{0pt}{4ex} 
\rule{0pt}{4ex} 

\end{tabular}\caption{Best-fitting parameter values for the fiducial double-Gaussian model for the samples used in our analysis. To account for sample variance we maximize the posterior probability function of the data for bootstrap samples of the clusters in each richness bin. The quoted uncertainties are the result of the model fitting over these bootstrap samples. We note that the uncertainties on these quantities measured on stacked cluster data is distinct from both the intrinsic variance in these quantities from cluster to cluster and the expected measurement uncertainty for these quantities for an individual cluster. }\label{tab:parameters} 
\end{center}
\end{table*}

\subsection{Luminosity dependence of projection effects}
\label{sec:results_lum_dep}

\begin{figure}
    \centering
    \includegraphics[width=0.5\textwidth]{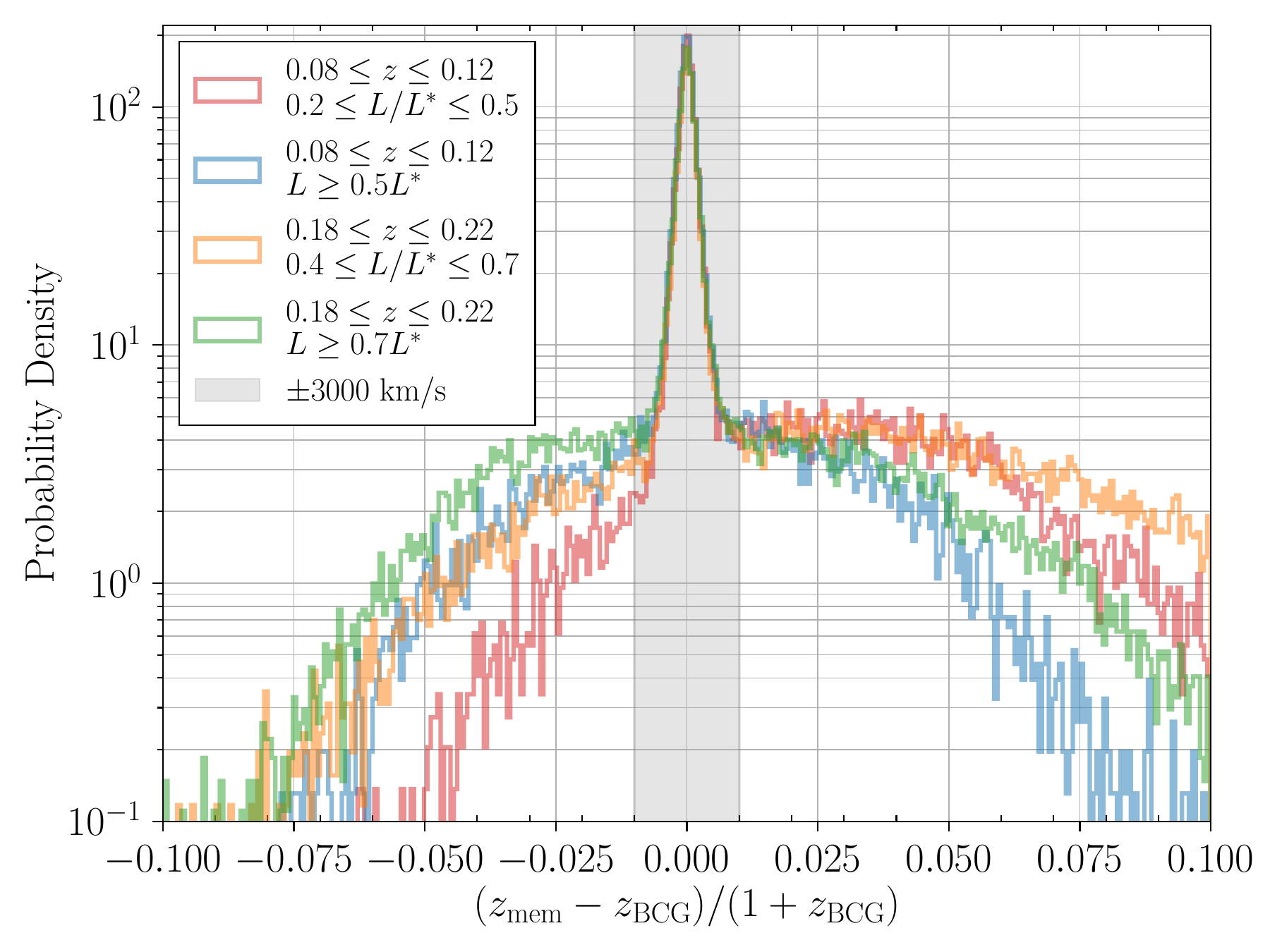}
    \caption[Luminosity dependence of projection effects]{Line-of-sight velocity distribution of \redmapper{} galaxy clusters for the low and high luminosity halves of the fiducial samples used in this analysis. At fixed redshift, the more extreme end of far-side projections is dominated by the fainter half of galaxies. The distribution of galaxies in projection at redshift $z\sim0.2$ also appears to have higher mean $\deltaznormed$. This illustrates that asymmetry in the distribution of projections is largely driven by faint galaxies.}
    \label{fig:v_los_dens_data_lum}
\end{figure}

\begin{figure*}
    \centering
    \includegraphics[width=0.9\textwidth]{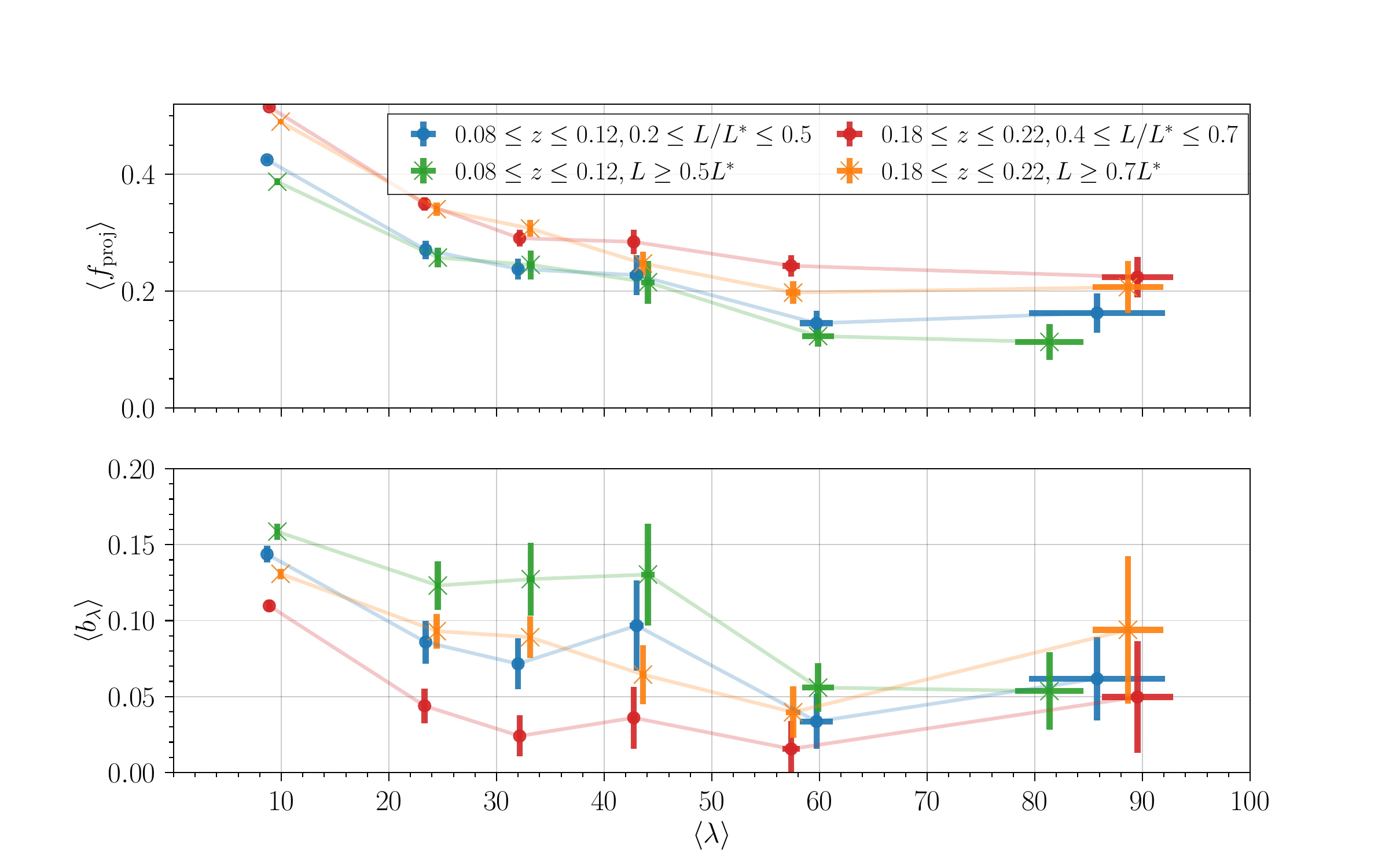}
    \caption[short]{Richness and redshift of projection effects for low and high luminosity halves of the fiducial samples in this work. Shown in the top panel is the value of the model parameter $\fproj$ encoding the fraction of putative \redmapper{} galaxies whose line-of-sight velocities indicate that these galaxies are \textit{not} true cluster halo members. The shape and relative amplitudes of the Gaussian mixture model components attributed to projection effects can vary for different samples; $\fproj$ summarizes the total amplitude of projection effects. This figure quantifies the luminosity dependence of projection effects: we find that at fixed redshift, the best estimate for the projection fraction is higher when using the fainter galaxy half for all but one (i.e., the $27.6 < \lambda \leq 37.6$) richness bin. 
    The richness bias, however, exhibits a partially reversed trend: because the \redmapper{} membership probabilities are lower on average for the fainter galaxy samples shown, the richness bias is lower when computed using the fainter galaxies, all else being fixed. Small $x$-axis offsets have been introduced to reduce overlap between points.
    }\label{fig:fproj_vs_lambda_lum}
\end{figure*}

\begin{figure}
    \centering
    \includegraphics[width=0.5\textwidth]{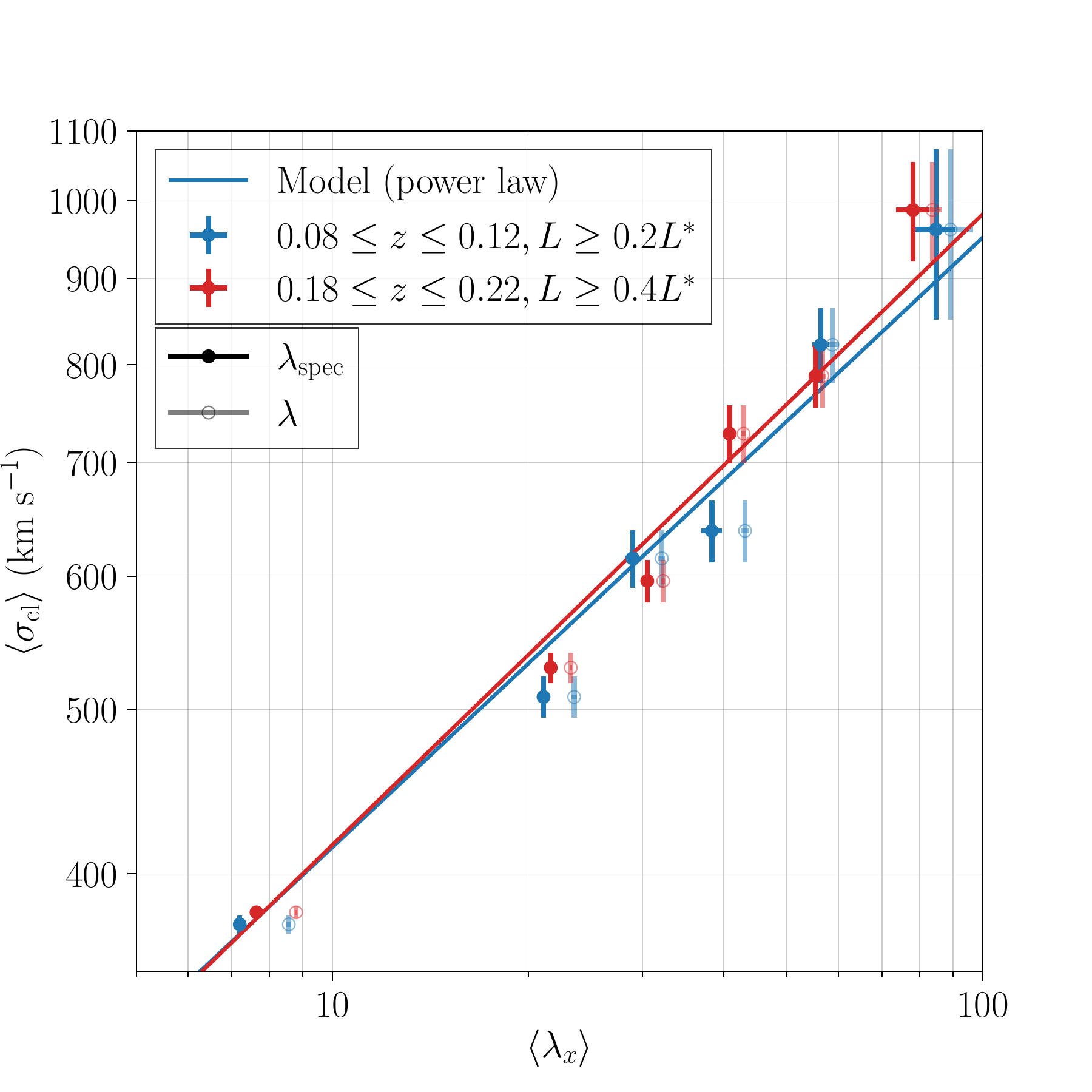}
    \caption[Cluster velocity dispersion vs. richness]{Stacked cluster velocity dispersion vs. richness for the fiducial samples in this work. Shown is the value of the model parameter $\sigmacl$ encoding the spread in velocities attributed to true cluster halo galaxies as a function of spectroscopic richness (filled markers, not transparent). Also shown for completeness are velocity dispersion as a function of \redmapper{} richness (open markers, transparent). While our line-of-sight velocity model provides only a probabilistic determination of whether any individual galaxy is a true cluster member given its line-of-sight velocity with respect to the central galaxy, $\sigmacl$ summarizes the cluster galaxy population  overall. Measured stacked velocity dispersion is largely similar at redshifts $z\sim 0.1$ and $z \sim 0.2$ as expected.
    The power-law fits of the trend in stacked velocity dispersion with spectroscopic richness are shown as colour curves and described in detail in \S \ref{sec:results_sigmacl}. Power law fits to the transparent curves are not shown for visual clarity, but are described in \S \ref{sec:results_sigmacl}. }\label{fig:sigmacl_vs_lambda}
\end{figure}

We find that the faintest galaxies in projection have redshifts at qualitatively more extreme line-of-sight velocities with respect to the main cluster halo, as illustrated in Figure \ref{fig:v_los_dens_data_lum}. Additionally, the fainter galaxies in projection appear more disproportionately on the far-side of the clusters under consideration, as the faint samples have projection distributions centred at a higher line-of-sight velocity than their brighter subsample counterparts. Asymmetry and possible non-Gaussianity in the \redmapper{} galaxy line-of-sight velocities with respect to their putative central galaxy emerges both for faint galaxies at $z\sim0.1$ and for galaxies at higher redshift ($z\sim0.2$). In this sense, projection effects are found in this work to be qualitatively luminosity dependent. The simple explanation for this is photometric noise: at fixed luminosity and cluster redshift, galaxies on the far side have noisier photometry and worse photometric redshifts than their counterparts on the near side. For a given cluster redshift, the far side will allow a larger range (relative to the near side) of intrinsic luminosities for which the corresponding observed magnitudes scatter into the red-sequence. This corresponds to a larger population of galaxies on the far side being possible projections, in addition to volume effects.

This manifests as the asymmetry illustrated comparing curves at fixed redshift in Figure \ref{fig:v_los_dens_data_lum}. However, how sensitively the \textit{fraction} of galaxies in projection and the richness bias depend on luminosity is quantitatively less striking. As shown in Figure \ref{fig:fproj_vs_lambda_lum}, the projection fraction and richness bias are consistent for all but the lowest richness bin. The faint half of the data demonstrates higher $\langle \fproj \rangle$ than its bright half counterpart for five of the six richness bins at both redshifts, but more data would be required to identify any statistically significant dependence for the individual richness bins.

Although the mean projection fraction increases for fainter galaxies, this is not the case for richness bias. In light of the lower \redmapper{} membership probabilities of fainter galaxies, fainter subsamples have lower richness bias on average than brighter subsamples. 

We note that a majority of low luminosity galaxies are relatively low $\pmem$ (59 and 66 per cent of $\pmem<0.2$ \redmapper{} galaxies satisfy $L \leq 0.50 L^*$ at $z \sim 0.1$ and $z \sim 0.2$, respectively). As such there is substantial overlap in the impact of the luminosity split shown here and imposing a conservative cut in $\pmem$ to construct richness.

\subsection{Velocity Dispersion and Spectroscopic Richness of \redmapper{} Galaxy Clusters}
\label{sec:results_sigmacl}

Our model for cluster galaxy line-of-sight velocities characterizes stacked velocity dispersion, which itself may prove a valuable cluster mass proxy. This velocity dispersion given by  parameter $\sigmacl$ in Equation \ref{eqn:halo_model} is shown as a function of spectroscopic richness in Figure \ref{fig:sigmacl_vs_lambda}. There is negligible cosmological evolution between these redshift ranges, so any discrepancy between velocity dispersions at the different redshifts shown here may indicate effects such as higher redshift clusters being more likely to be dynamically unrelaxed or that at fixed richness \redmapper{} is selecting clusters of different physical halo mass at different redshifts. Given uncertainties, the velocity dispersion of the fiducial and conservative subsamples are consistent, lending credence to the robustness of the model.

We fit a power-law $\sigmacl = a \lambdatrue^k$ to the observed relation between $\sigmacl$ and $\lambdatrue$, finding best-fitting slopes of $k_{z\sim0.1}=0.36 \pm 0.03$ and $k_{z\sim0.2} = 0.37 \pm 0.03$ at $z\sim0.1$ and $z\sim0.2$, respectively. This is lower than the $0.44 \pm 0.02$ slope for \redmapper{} clusters quoted in \citet{redmapper4}, a difference we attribute to our projection effect model more accurately accounting for the velocity dispersion of true cluster galaxies and that of galaxies in projection. When fitting the power law model using mean \redmapper{} richness instead of spectroscopic richness, we obtain $k_{z\sim0.1}=0.38 \pm 0.03$ and $k_{z\sim0.2} = 0.39 \pm 0.03$ at $z\sim0.1$ and $z\sim0.2$, respectively. This suggests that the difference in velocity dispersion definitions is more important than the difference in photometric vs.~spectroscopic richness in causing the discrepancy we observe relative to \citet{redmapper4}. Using hydrodynamical simulations, \citet{Munari} predict a slope for the relation between cluster galaxy velocity dispersion and halo mass for virialized systems, $\sigmacl \propto M^{0.364\pm0.01}$ (where the systematic uncertainty on the index varies depending on the astrophysical feedback implementation). 
Together, these relations confirm an implied approximately linear relationship between $\lambdatrue$ and the three-dimensional halo mass, with $\lambdatrue$ scaling as mass to the power $\alpha_{z\sim0.1} = 1.01 \pm 0.09$ and $\alpha_{z\sim0.2} = 0.98 \pm 0.08$.
We note that this slope fit to the $\sigmacl$ to $\lambdatrue$ relation is highly constrained by the low richness bin, which is by far the largest sample by number of clusters and galaxies. The uncertainty decreases for larger samples, consistent with being interpretable as an uncertainty on the mean quantity fit to the stacked data.  Since the slope found is dependent on this bin, however, we highlight that the goodness-of-fit metrics of $\chi^2_{\rm red}=2.6$ ($p=0.04$) at $z\sim0.1$ and $\chi^2_{\rm red}=3.3$ ($p=0.01$) at $z\sim0.2$ suggest the uncertainties reported may likely be underestimated.

\section{Robustness Tests}
\label{sec:robustness_tests}

\begin{figure}
  \includegraphics[width=0.5\textwidth]{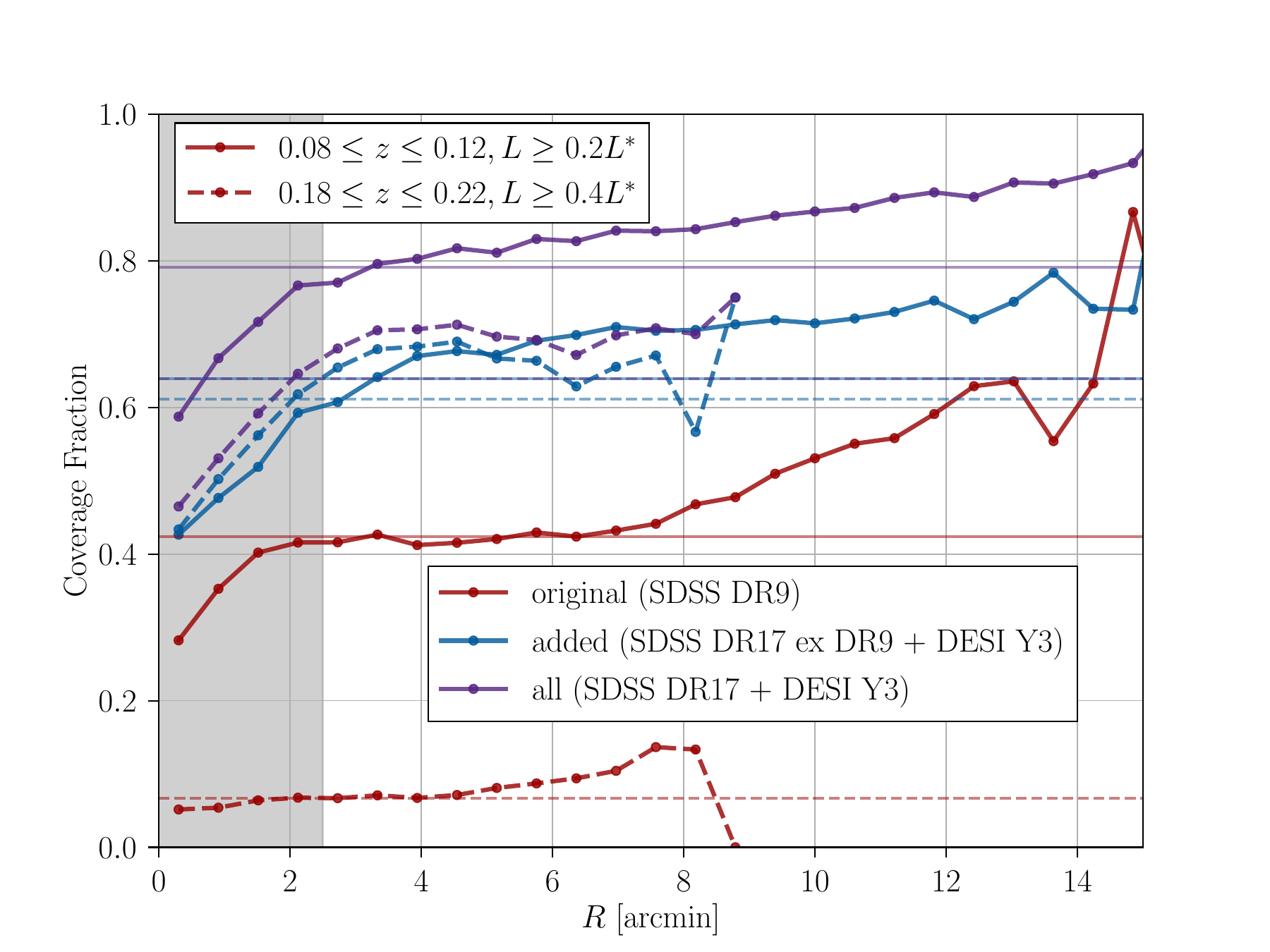}
  \caption[DESI fibre collisions]{Fraction of \redmapper{} cluster galaxies with spectroscopic redshifts (accounting for footprint differences) using the spectroscopic data in this study at $z\sim0.1$ (solid) and $z\sim0.2$ (dashed), respectively. Fibre collisions are known to limit coverage below $2\farcm{4}$ (grey region); the DESI selection additionally shows evidence of radial dependence at higher angular distances.  We demonstrate robustness to this selection effect on the data with an alternative analysis described in \S \ref{sec:robustness_tests_fiber}.}
  \label{fig:fiber_collision_plot}
\end{figure}

\begin{figure*}
    \centering
    \includegraphics[width=0.75\textwidth]{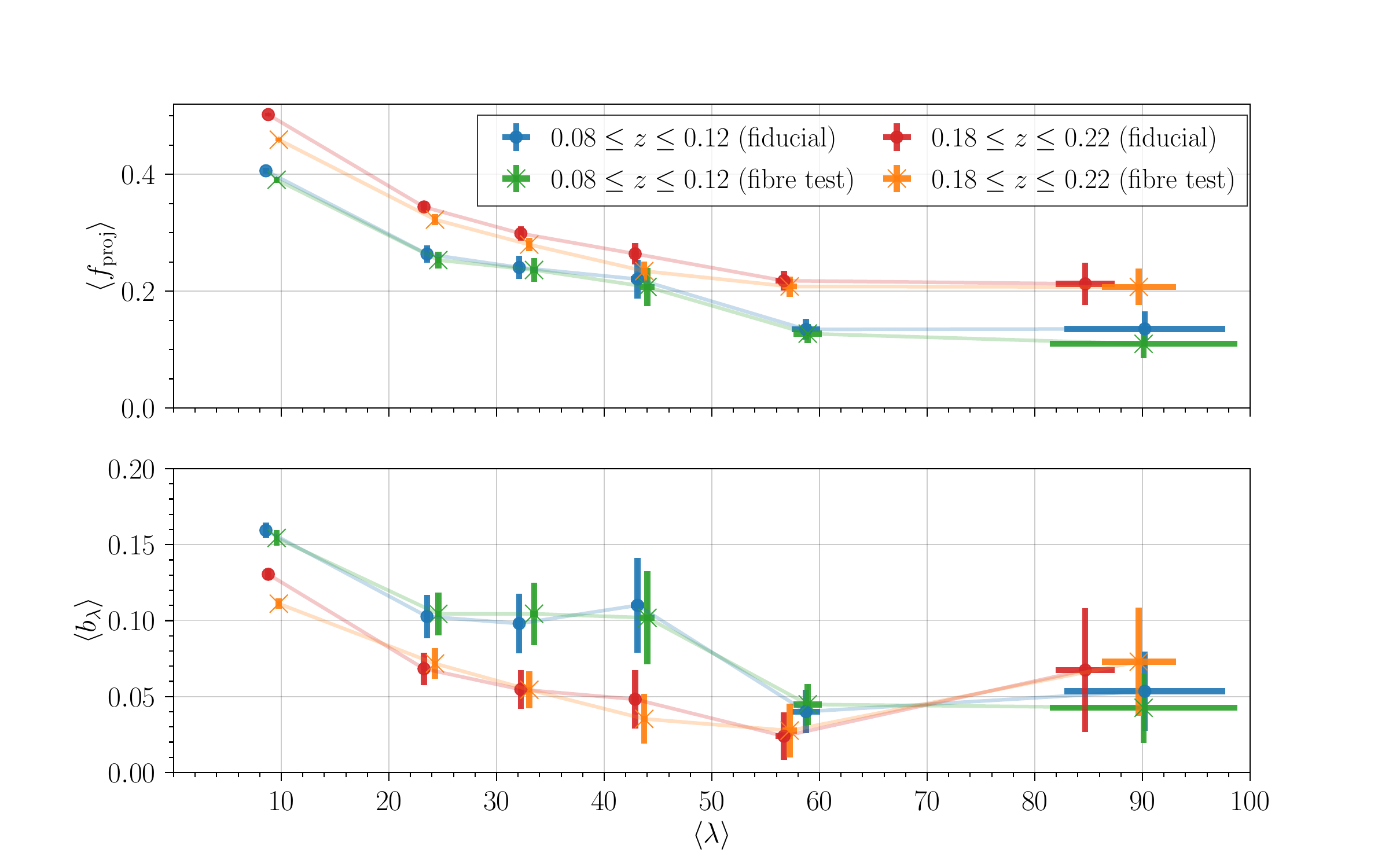}
    \caption[Robustness of results to DESI fibre collisions]{Richness and redshift dependence of projection effects for the fiducial and alternative analyses illustrating robustness of our results to impacts of DESI fibre collisions. Markers with diagonal crosses indicate the alternative analysis, while the single vertical cross markers illustrate the fiducial analysis. In the alternative analysis galaxies without observed spectroscopic redshifts are assigned the redshift of the nearest neighboring \redmapper{} cluster galaxy that has a spectroscopic redshift (up to the fifth nearest neighbor). This has the effect of up-weighting redshifts in dense cluster cores, thus counteracting their lower sampling due to observational constraints. The general agreement between these constraints for all but one richness bin suggests our results are not sensitive to the impact of DESI fibre collisions on the sample. Small $x$-axis offsets have been introduced to reduce overlap between points.\label{fig:fproj_vs_lambda_fib}
    }
\end{figure*}

\begin{figure*}
\label{fig:fproj_vs_lambda_color}
    \centering
    \includegraphics[width=0.75\textwidth]{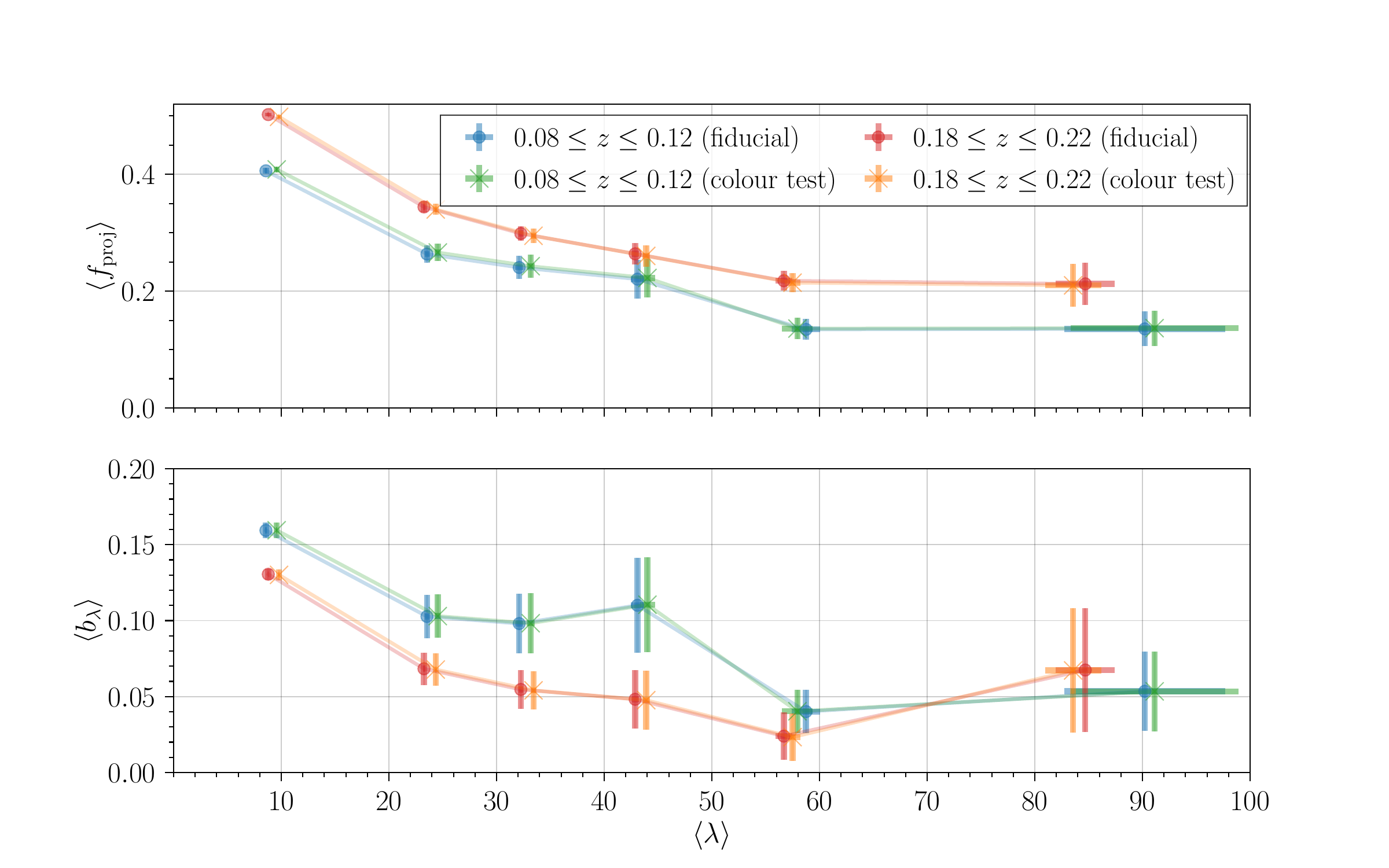}
    \caption[Robustness to colour selection]{Richness and redshift dependence of projection effects for the fiducial and alternative analyses illustrating robustness of our results to impacts of colour dependence of the DESI selection functions. Markers with diagonal crosses indicate the alternative analysis, while the single vertical cross markers illustrate the fiducial analysis. In the alternative analysis the importance of each galaxy in the likelihood is weighted such that the weighted distribution in $g-r$ colour of the cluster galaxy sample with spectroscopic redshifts matches the corresponding colour distribution of the parent cluster galaxy population. The general agreement between these constraints suggests our results are not sensitive to the impact of the colour dependence of the DESI selection functions. In practice, the colour distributions of the photometric and spectroscopic samples are similar, meaning the weights are largely approximately $\sim1$. As a result the two runs illustrated here are highly correlated. Small $x$-axis offsets have been introduced to reduce overlap between points.
    }
\end{figure*}

Here we present two alternative analyses to demonstrate the robustness of our results to the residual impact of the DESI spectroscopic selection on the sample of cluster galaxies used to measure velocity dispersion and projection effects. Ultimately, if the probability of DESI to yield a spectroscopic redshift for a \redmapper{} cluster galaxy is correlated with the probability of that galaxy being a true member, the measurements presented in \S \ref{sec:results} would have an associated systematic error. 

\subsection{Robustness to DESI fibre collisions}
\label{sec:robustness_tests_fiber}

Not all cluster galaxies satisfying the selection criteria in this work (cluster redshift and richness as well as galaxy luminosity) have a spectroscopic redshift, with DESI fibre collisions being a major limiting factor preventing observation of galaxies in the dense cores of clusters. The minimum pitch between neighboring fibres is 10.4mm. Given the average focal plane plate scale of 70.8 $\mu$m$/\arcsec$ this corresponds to $\sim2\farcm{4}$ . The \redmapper{} cluster radius of the typical $(\lambda\sim30)$ cluster $(R_\lambda \sim 0.79 \ h^{-1} \text{ Mpc}$) corresponds to $5\arcmin$ and 2\farcm{75} at redshifts $z\sim0.1$ and $z\sim0.2$, respectively. This does not allow for significant sampling of an individual typical cluster, with the fraction of members with spectra improving (deteriorating) for higher (lower) richness than the typical $\lambda\sim30$ value quoted here.

As shown in Figure \ref{fig:fiber_collision_plot}, DESI coverage steadily falls as the projected distance from the central galaxy decreases. At $z\sim0.1$, the coverage fraction including DESI data rises to near unity as distance from the central galaxy increases. At $z\sim0.2$ the coverage fraction levels off at the average coverage of roughly 60 per cent.  This introduces a selection effect whereby galaxies farther in projected radial distance from the cluster centre are more likely to be observed and included in our analysis. If the probability of being in projection correlates with this distance, our constraints on $\fproj$ and $b_\lambda$ would have a corresponding systematic bias. 

To assess the degree of this potential systematic uncertainty, we conduct an alternative analysis in which galaxies in the \redmapper{} member catalogue with no spectroscopic data are assigned mock spectroscopic redshift values. In particular, galaxies without real observed spectroscopic redshifts are assigned the redshift of the nearest neighboring \redmapper{} cluster galaxy that has a spectroscopic redshift (considering as far in angular separation as the fifth nearest neighbor). Galaxies with real observed spectroscopic redshifts retain those values from the fiducial analysis. 

Since the projected cluster galaxy density profile falls as radius from the cluster centre increases, the effect of this choice is to up-weight the importance of spectroscopic redshifts in the cluster cores. Such redshifts will be repeated more often than spectroscopic redshifts in the cluster outskirts. If these galaxies had higher (lower) probability of being in projection relative to their outskirt counterparts, we would observe an increase (decrease) in projection effects in this alternative analysis.

As shown in Figure \ref{fig:fproj_vs_lambda_fib}, we find our results in this alternative analysis to be consistent with those of our fiducial analyses for all richness-redshift bins except the low richness bin $5 \leq \lambda \leq 20$ at $z\sim0.2$. Even for the sole discrepant richness-redshift bin, the main findings of richness and redshift dependence of projection effects hold. We therefore conclude that this robustness test indicates that the constraints on projection effects described in this work do not depend sensitively on fibre collision effects. We note that $\fproj$ is almost always lower in the fibre test case compared to to the fiducial, but this trend is largely eliminated in $b_\lambda$. This is consistent with the plausible notion that galaxies in the projected centre of a cluster are more likely to be cluster members, but since \redmapper{} accounts for this spatial dependence in its membership probability the richness bias is more consistent between the runs.

\subsection{Robustness to DESI colour selection}
\label{sec:robustness_color}
The cluster galaxy sample analysed here is 
subject to the colour-dependence of the DESI sample selection. Our results are therefore dependent on any correlation between the probability that a \redmapper{} galaxy is in projection and the probability of having a spectroscopic redshift for that galaxy, after the sample selection criteria of \S \ref{sec:data} are applied to the data. We conduct an additional alternative analysis to account for colour-dependence of the DESI selection. 

In this an alternative analysis each galaxy in the \redmapper{} member catalogue with a measured spectroscopic redshift is assigned a weight. The weight is designed so that the distribution in $g-r$ colour of the subsample of galaxies with spectra matches that of the distribution of the full sample of \redmapper galaxies (i.e. including galaxies without spectroscopic redshifts). The weight is set by the ratio of the normalized histograms in $g-r$ colour of the photometric and spectroscopic samples. Any change in the results of the primary measured model parameters ($\fproj$ and $\sigmacl$) reflects a systematic uncertainty due to impact of the DESI colour selection function (both the explicit galaxy targeting function and the implicit selections such as redshift success) on the spectroscopic sample available for characterizing the clusters.

The results, shown in Figure \ref{fig:fproj_vs_lambda_color} are virtually identical to their counterparts in the fiducial analyses. This indicates that the model constraints for projection effects shown in this work are not sensitively dependent on the colour-dependence of the DESI spectroscopic selection at $z \lesssim 0.2$.

\section{Conclusions}
\label{sec:conclusions}

Our results confirm the main findings of \citetalias{Myles2021}, namely that projection effects are substantial and richness dependent, while spectroscopic richness and stacked velocity dispersion are promising mass proxies for galaxy cluster cosmology. Our approach using spectroscopic data is complementary to calibrating projection effects with photometric information \citep[e.g.,][]{Costanzi2019}.

We build upon the work of \citetalias{Myles2021} in finding the first evidence of redshift dependence in the observed signal, consistent with the interpretation that excess projections are driven by photometric redshift uncertainties and the width of the red sequence. This redshift dependence is cause for concern. While the severity of the problem in question is expected to increase at higher redshifts, the fact that it does so substantially at such a small increment in redshift heightens the risks associated with using projection effect models constrained non-empirically. The 4000-\AA{} break falls in the same photometric filter ($g$) at both of the redshifts analysed in this work. This leaves the presumable causes of the increase to be increased photometric noise in the same filters, decreased spectroscopic training data of the red-sequence, and the intrinsic width of the red sequence. 

Taking the results into account, we conclude additional spectroscopic data may be needed to calibrate the \redmapper{} red-sequence model for cosmological applications. This also motivates additional observations: a survey of a representative set of clusters in bins of richness and redshift across the full redshift range of the expected Rubin LSST sample (out to $z\sim1$ for red-sequence based cluster samples identified using Rubin filters alone, \citealt{lsst_sci_book}). This would enable empirical constraints on projection effects for LSST clusters. Understanding any further increases and/or a plateau in the trend of projection effects with redshift is essential for calibrating the mass-richness relation. 

Finally, we also find that projection effects are qualitatively luminosity dependent, with the faintest putative cluster galaxies ($0.2 \leq L/\lstar \leq 0.55$) exhibiting stronger non-Gaussian tails in line-of-sight velocity. While the faintest galaxies contribute relatively little to the amplitude of projection effects, as noted by the consistency in $\fproj$, the extreme projections are disproportionately drawn from this faint end, with potentially meaningful implications for the scatter of individual objects with low true richness into bins with high observed richness. 

A dedicated observing campaign spanning the relevant redshift and richness ranges ($\lambda \gtrsim 20$, $0 \leq z \lesssim 1$) is essential to adequately calibrating projection effects for optical cluster cosmology. While costly to secure redshifts for galaxies as faint as these in this redshift range, such a survey would enable a proper measurement at the most cosmologically valuable redshift of LSST.

\section{Discussion}
\label{sec:discussion}

We speculate on the implication of our results of richness and redshift dependence in projection effects for previous cosmology findings. These results have implications on the interpretation of \redmapper{} cluster weak lensing measurements such as those used in \citet{desy1}. Figure \ref{fig:fproj_vs_lambda} shows that, for low-to-intermediate richness systems, on average, a substantial fraction of the galaxies identified by \redmapper as being associated with a cluster halo will be line-of-sight projections. The mass associated with these projected galaxies will contribute to the observed lensing signals. However, the lower mass-to-light ratios of field galaxies compared to cluster members ($\sim5-10\times$; \citealt{Dai0911.2230}) will lead to these lensing boosts being small (roughly a few per cent). The net result is that, for the richest clusters, both richness and weak lensing mass should be measured relatively accurately. By contrast, for the least rich systems, projection effects will cause richness to be biased high and the implied lensing mass at a given richness to be biased low. The median cluster redshift of SDSS \redmapper{} clusters with $\lambda \geq 20$ is $z\sim0.5$ (and higher in DES and LSST). As a result, the degree to which this bias affects the majority of the clusters in recent cosmology analyses depends on the degree of change in projection fraction beyond redshift $z\sim0.2$ relative to the change in average \redmapper{} membership probabilities. 

The boost in richness due to projection effects can bias the observed cluster halo number density high. In this context, our results may be relevant for the interpretation of recent cosmology analyses based on cluster observations with \erosita{} \citep{erosita1, erosita_cosmo}. In their analysis, galaxy clusters were detected via X-ray emission from the intracluster medium. To clean the sample of X-ray point sources (namely, AGN) that can be confused for extended emission due to the point spread function, the cosmology sample used only those X-ray detections confirmed with a \redmapper-based algorithm. The threshold in optical richness of $\lambda > 3$  \citep{erosita_cosmo, Kluge2024} used for this purpose, however, may be insufficient to achieve the goal of removing false positive \erosita{} cluster detections. As shown in Figure \ref{fig:fproj_vs_lambda}, on average over half of the galaxies associated with such low richness detections will be projections. As a result, one can expect that many \redmapper{} detections satisfying $\lambda > 3$ are false positive cluster detections. The corresponding result would be an overestimated cluster number density, and a correspondingly higher value in $S_8$. Based on this reasoning, we recommend a higher richness threshold for confirming ICM-based cluster detections, though the exact optimal value will depend on constraints of the full relationship $p(\lambdatrue | \lambda)$ (as opposed to the mean constrained in this work). A reduction in $S_8$ would bring the \erosita{} results ($S_8 = 0.86 \pm 0.01$) closer to the results from \planck{} as well as other cluster cosmology constraints \citep{erosita_cosmo}. 

We also consider our results in the context of past work modeling the impact of projection effects in \redmapper{} galaxy clusters. \citet{Lee2024} used a Halo Occupation Distribution (HOD) to populate simulated dark matter haloes with red-sequence galaxies to model the impact of projection effects on cluster observables. Measuring the impact of projection effects on richness with a counts-in-cylinders metric, they find their model prediction for $\fproj$ is consistent with the empirical constraints from \citetalias{Myles2021} at $z\sim0.1$. Our results showing higher $\fproj$ at $z\sim0.2$ than at $z\sim0.1$ can naturally be interpreted as corresponding to a larger characteristic length scale on which galaxies in projection are selected as cluster galaxy candidates by \redmapper{}. Our empirical constraint may motivate redshift-dependent parameters for the modeling framework developed by \citet{Lee2024}. In a separate analysis, \citet{Sunayama2023} measured the impact of projection effects on the cluster lensing and cluster clustering signals, finding anisotropic boosts in these observables due to preferential selection of filaments aligned with the line-of-sight as a cluster. While $\fproj$ is not a direct parameter of the projection effect model in \citet{Sunayama2023}, our results suggest there may be benefits to incorporating into this model a spectroscopically derived constraint on $d_{\mathrm{proj}}$ (i.e., the distance along the line of sight within which galaxies in projection are misidentified as cluster members).

Having confirmed the main findings of \citetalias{Myles2021} and used the more numerous, fainter, and higher redshift sample available with DESI to identify redshift- and luminosity- dependence of projection effects, several avenues of future work on characterizing galaxy clusters with representative spectroscopy become clear. We enumerate several of these directions of future work below:

\begin{enumerate}
    \item \textbf{Targeted spectroscopic follow-up of \redmapper{} clusters in bins of redshift and richness:}
    These results motivate collecting additional data at all cosmologically relevant redshifts to measure velocity dispersion and projection effects. While our result that projection effects become more severe at higher redshifts is not unexpected, the relatively steep increase (by as much as 25 per cent) for such a modest increase in redshift ($\delta z \sim 0.1$) heightens the importance of empirical constraints. Assuming the trend persists for fainter galaxies it will be essential to calibrate projection effects with representative spectroscopy to make use of any cluster sample selected via Rubin photometry. 

    \item {\textbf{Measure individual cluster velocity dispersion and spectroscopic richness}: }
    Targeted follow-up data of \redmapper{} clusters would facilitate measuring velocity dispersion and spectroscopic richness for individual clusters. Conducting this analysis on individual clusters and combining the constraints on relevant model parameters from the individual fits would constrain the scatter on these parameters rather than simply the uncertainty on the means. Importantly, this would constrain the full distribution $p(\lambdatrue|\lambdaobs)$.

  \item {\textbf{Improving the \redmapper{} algorithm}:}
    Our results suggest various lines of reasoning to improve \redmapper{}. 
    First, we recommend re-training the \redmapper{} red sequence with DESI data to investigate any potential differences in richness estimates resulting from the addition of this spectroscopic training data.
    Second, we recommend exploring further tuning of the lower luminosity threshold used by \redmapper{}. Similarly, the lower threshold in $\pmem$ used for constructing richness merits further investigation.
    Moreover, incorporating a spectroscopic matched-filter into the \redmapper{} algorithm itself could provide the most robust results in the regime in which the number of galaxy spectra is being significantly increased with new experiments. 

  \item {\textbf{Improved model for putative cluster galaxy velocities:}} The addition of fainter ($L<0.55 \lstar$) and higher redshift $z \sim 0.2$ cluster galaxies illustrates deviation from a zero-centred Gaussian in the distribution of line-of-sight velocities of projection effects. In this study, this complication was evaded by truncating extreme projections at $|\deltaznormed|>0.1$. However, it is known in principle that the observed distribution has contributions from true cluster galaxies, infalling galaxies, galaxies associated with clusters whose centres are mis-identified, and projections in correlated and uncorrelated structure along the line-of-sight. Moreover, neighboring cluster pairs along the line-of-sight can be confused for single detections. As the amount of data increases, so too may the complexity of the model necessary to fit the data. One could also optionally adopt a luminosity threshold that smoothly varies with redshift to further probe redshift dependence of projection effects.

We emphasize that the ultimate goal of measuring an unbiased, low-scatter proxy for cluster mass allows for flexibility in sacrificing model interpretability. A more general Gaussian mixture model may achieve this goal. Our current model has the distinct advantage of a simple interpretation: one component for true cluster galaxies and one for all other galaxies, with a clear probability for any given galaxy to be one or the other depending solely on its line-of-sight velocity. We note that many modern statistical and machine learning methods could explore the space of models that relate halo mass to the full distribution of galaxy line-of-sight velocities.

  \item {\textbf{Combining velocity dispersion and weak lensing mass to study cosmological and gravitational physics:} Galaxy cluster velocity dispersion is a complementary mass proxy to weak lensing. It has unique value as a \textit{dynamical} mass proxy that is observationally feasible to measure across the parameter space of cluster mass and redshift. A cluster count cosmology analysis with the mass-observable relation given by $p(M|\lambda) \propto p(\lambda|v_{\text{cl}}, \text{WL})$ would benefit from the complementarity of lensing and dynamical masses. 
    Moreover, by comparing a \textit{dynamical} cluster mass proxy sensitive to the Newtonian potential (velocity dispersion) with cluster lensing mass sensitive to the full gravitational metric potential, we can constrain the `gravitational slip' quantifying deviation from general relativity \citep{Joyce2016, Pizutti2019}. The availability of spectroscopic mass proxies down to the galaxy group scale across a broad range of redshifts may prove this approach as competitive for testing gravitational physics. While the use of cluster velocity dispersion as a mass proxy would require sufficient constraints on galaxy velocity bias from simulations, recent results suggesting per cent level uncertainty in cluster galaxy velocity bias lend credence to the potential feasibility of this approach \citep[see e.g., ][]{Evrard2008, Anbajagane2022}.}
\end{enumerate}

\section*{Acknowledgements}
The authors would like to thank Eric Huff, Yen-Ting Lin, Peter Melchior, Eduardo Rozo, and Michael Strauss for highly valued scientific discussion. 

We thank the internal reviewers in the DESI collaboration and the journal referee for their many helpful comments on this work. 

Support for this work was provided by The Brinson Foundation through a Brinson Prize Fellowship grant.

Funding for the Sloan Digital Sky Survey IV has been provided by the Alfred P. Sloan Foundation, the U.S. Department of Energy Office of Science, and the Participating Institutions. SDSS-IV acknowledges
support and resources from the Center for High-Performance Computing at
the University of Utah. The SDSS web site is www.sdss.org.

SDSS-IV is managed by the Astrophysical Research Consortium for the 
Participating Institutions of the SDSS Collaboration including the 
Brazilian Participation Group, the Carnegie Institution for Science, 
Carnegie Mellon University, the Chilean Participation Group, the French Participation Group, Harvard-Smithsonian Center for Astrophysics, 
Instituto de Astrof\'isica de Canarias, The Johns Hopkins University, Kavli Institute for the Physics and Mathematics of the Universe (IPMU) / 
University of Tokyo, the Korean Participation Group, Lawrence Berkeley National Laboratory, 
Leibniz Institut f\"ur Astrophysik Potsdam (AIP),  
Max-Planck-Institut f\"ur Astronomie (MPIA Heidelberg), 
Max-Planck-Institut f\"ur Astrophysik (MPA Garching), 
Max-Planck-Institut f\"ur Extraterrestrische Physik (MPE), 
National Astronomical Observatories of China, New Mexico State University, 
New York University, University of Notre Dame, 
Observat\'ario Nacional / MCTI, The Ohio State University, 
Pennsylvania State University, Shanghai Astronomical Observatory, 
United Kingdom Participation Group,
Universidad Nacional Aut\'onoma de M\'exico, University of Arizona, 
University of Colorado Boulder, University of Oxford, University of Portsmouth, 
University of Utah, University of Virginia, University of Washington, University of Wisconsin, 
Vanderbilt University, and Yale University.

This research used data obtained with the Dark Energy Spectroscopic Instrument (DESI). DESI construction and operations is managed by the Lawrence Berkeley National Laboratory. This material is based upon work supported by the U.S. Department of Energy, Office of Science, Office of High-Energy Physics, under Contract No. DE–AC02–05CH11231, and by the National Energy Research Scientific Computing Center, a DOE Office of Science User Facility under the same contract. Additional support for DESI was provided by the U.S. National Science Foundation (NSF), Division of Astronomical Sciences under Contract No. AST-0950945 to the NSF’s National Optical-Infrared Astronomy Research Laboratory; the Science and Technology Facilities Council of the United Kingdom; the Gordon and Betty Moore Foundation; the Heising-Simons Foundation; the French Alternative Energies and Atomic Energy Commission (CEA); the National Council of Humanities, Science and Technology of Mexico (CONAHCYT); the Ministry of Science and Innovation of Spain (MICINN), and by the DESI Member Institutions: www.desi.lbl.gov/collaborating-institutions. The DESI collaboration is honored to be permitted to conduct scientific research on I’oligam Du’ag (Kitt Peak), a mountain with particular significance to the Tohono O’odham Nation. Any opinions, findings, and conclusions or recommendations expressed in this material are those of the author(s) and do not necessarily reflect the views of the U.S. National Science Foundation, the U.S. Department of Energy, or any of the listed funding agencies.

This research used resources of the National Energy Research Scientific Computing Center (NERSC), a Department of Energy Office of Science User Facility

This work made use of software cited as follows: \citet{numpy, scipy, scikit-learn, emcee, pandas, jax, Wright2006}


\section*{Data Availability}
This paper made use of the DESI DR2 spectroscopic catalog, which will be made public per the DESI data policy. 

Data points shown in figures are available in machine-readable format at \url{https://doi.org/10.5281/zenodo.15589629}.



\bibliographystyle{mnras}
\bibliography{references, desi, oss} 




\appendix

\section{Validation of fiducial line-of-sight velocity model}
\label{sec:model_validation}
In this appendix we demonstrate parameter posteriors for a fit of the model to the data with the cluster mean line-of-sight velocity as a free parameter and the component of the model representing galaxies in projection fit independently to each richness bin. 

\begin{figure*}
  \includegraphics[width=0.45\textwidth]{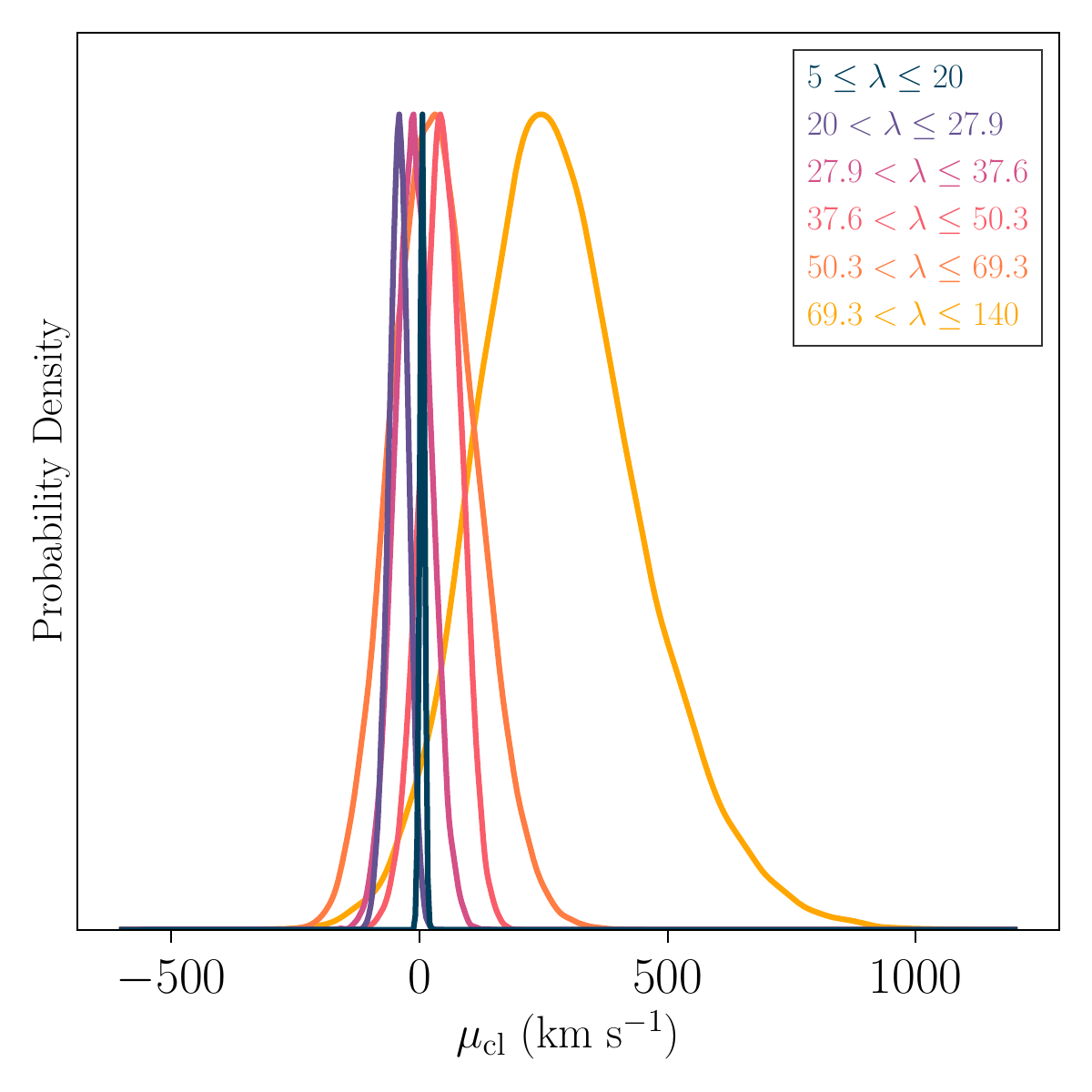}%
  \hfill
  \includegraphics[width=0.45\textwidth]{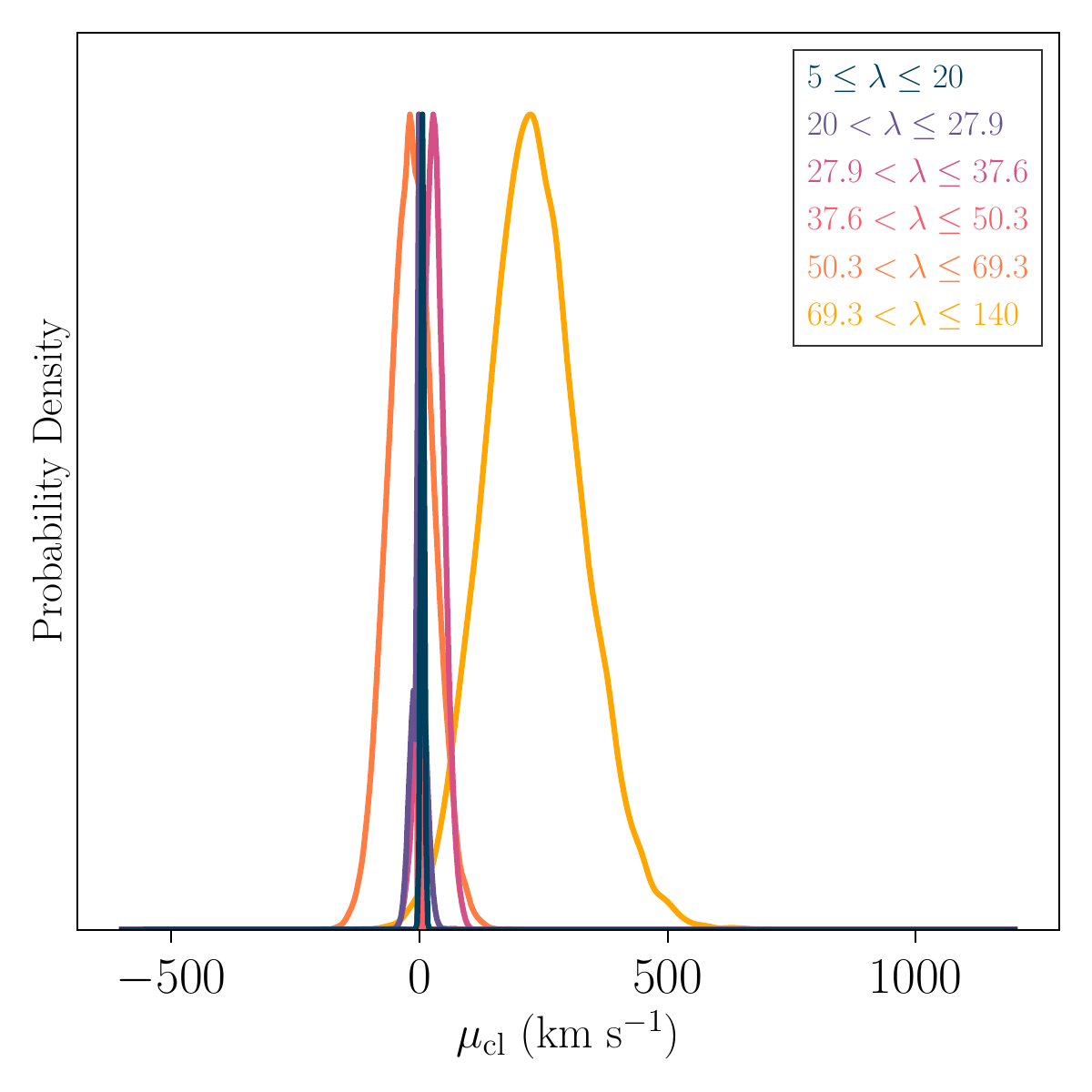}
  \caption[Cluster component mean posterior]{Cluster component mean line-of-sight velocity for model fits where this parameter is free and the model is fit independently for each richness bin at $z\sim0.1$ (left) and $z\sim0.2$ (right), respectively. The broad consistency of the richness bins (excepting slight tension for the highest richness bin discussed in Appendix \ref{sec:model_validation}) is used to justify fixing the cluster component mean to zero in the fiducial model fits.}
  \label{fig:cluster_mu_assumptions}
\end{figure*}

Figure \ref{fig:cluster_mu_assumptions} illustrates the justification for fixing the cluster component mean to zero line-of-sight velocity. At $z\sim0.1$, all richness bin are empirically consistent with zero mean line-of-sight velocity; at $z\sim0.2$ all but the highest richness bin remain consistent, with the highest richness bin exhibiting some tension with mean zero. This tension in the highest richness bin may indicate an overall imperfection in the model, e.g., the constraint on the cluster component of the model in Equation \ref{eqn:halo_model} may be biased by the projected component in the regime of smaller sample size. We nevertheless choose to fix the cluster mean line-of-sight velocity to zero for the fiducial model at both redshifts due to the expected common cause for projection effects: photometric redshift uncertainty and the width of the red sequence. 

\begin{figure*}
  \includegraphics[width=0.45\textwidth]{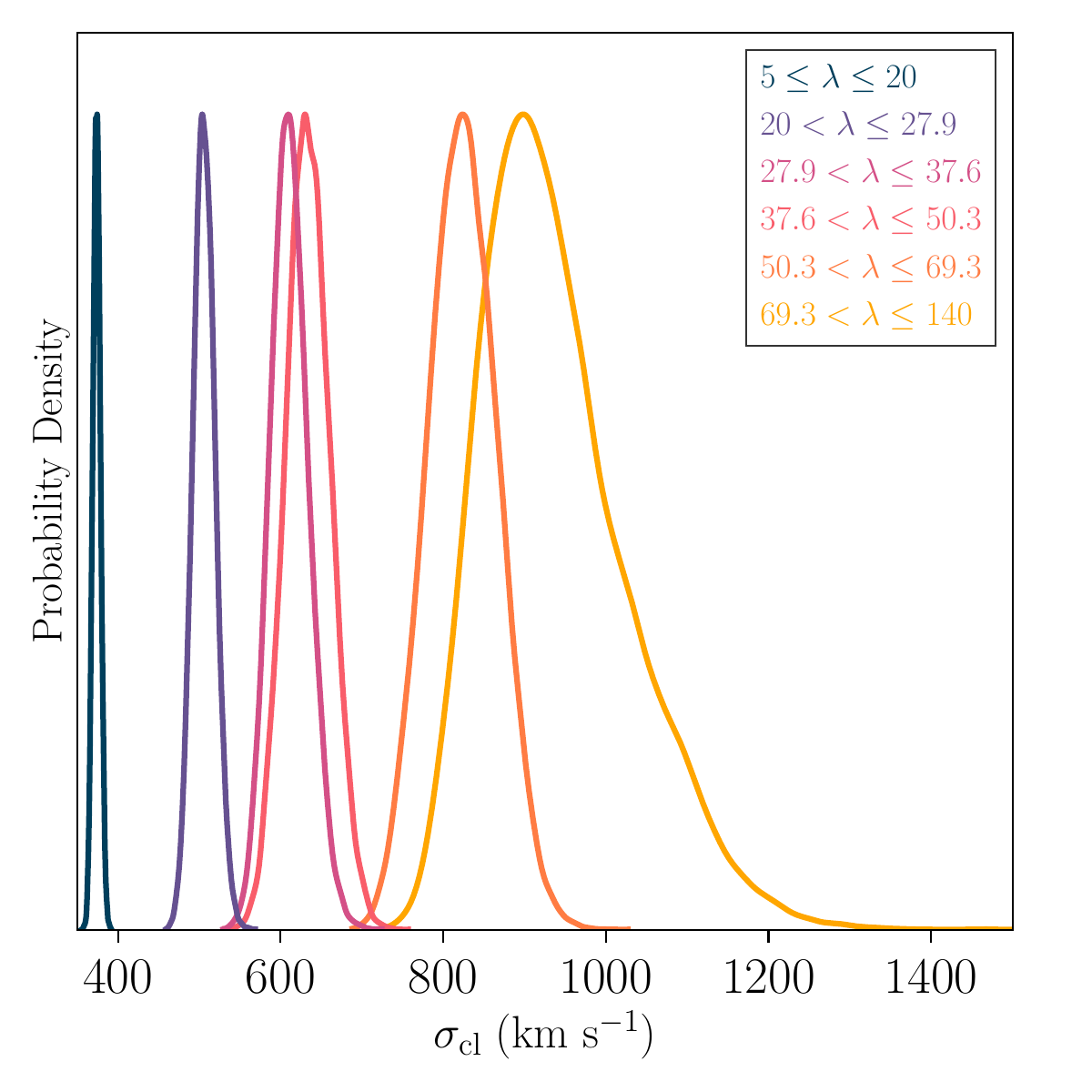}%
  \hfill
  \includegraphics[width=0.45\textwidth]{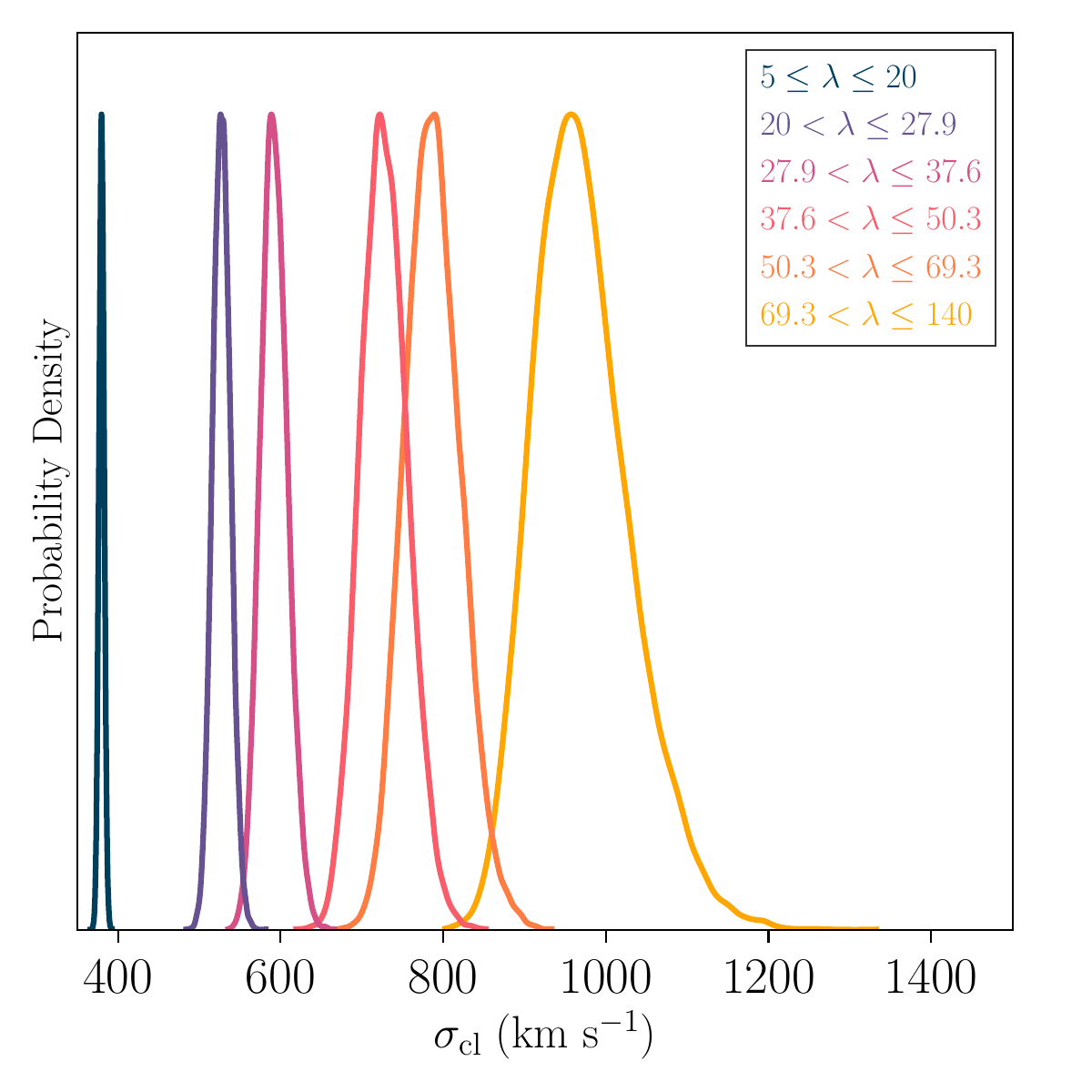}
  \caption[Cluster velocity dispersion posterior]{Cluster component velocity dispersion for model fits where the mean line-of-sight parameter is free and the model is fit independently for each richness bin at $z\sim0.1$ (left) and $z\sim0.2$ (right), respectively.}
  \label{fig:cluster_sig_assumptions}
\end{figure*}

Figure \ref{fig:cluster_sig_assumptions} demonstrates the posterior on velocity dispersion in this free model fit. The tail of the velocity dispersion of the highest richness bin may indicate confusion of the cluster and projected component for this subsample. 

\begin{figure*}
  \includegraphics[width=0.45\textwidth]{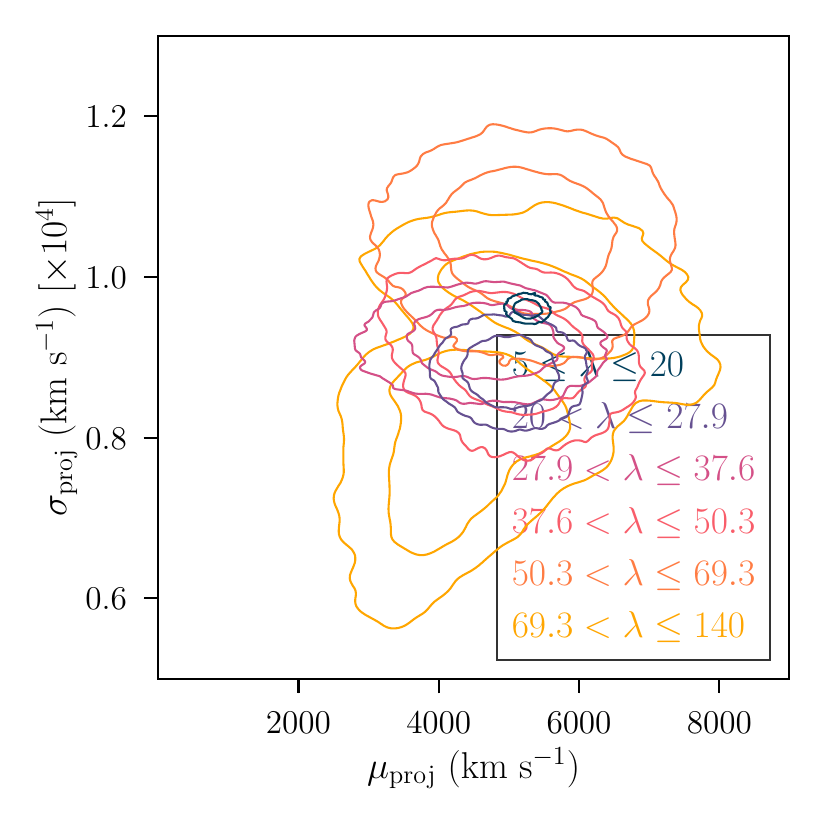}%
  \hfill
  \includegraphics[width=0.45\textwidth]{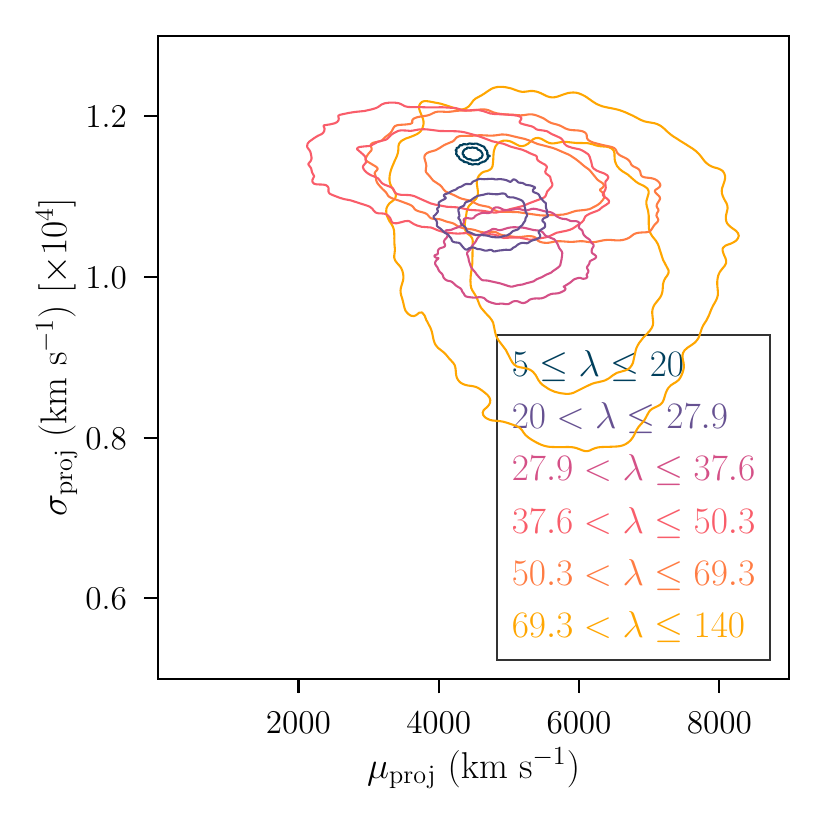}
  \caption[Projection component posterior]{Projection component mean line-of-sight velocity and velocity dispersion for model fits where the mean parameter is free and the model is fit independently for each richness bin at $z\sim0.1$ (left) and $z\sim0.2$ (right), respectively. We make the approximation that the consistency between richness bins is sufficient to fit the projection component jointly to data from all richness bins.}
  \label{fig:proj_assumptions}
\end{figure*}

Figure \ref{fig:proj_assumptions} demonstrates the projection component posterior when it is fit independently to each richness bin. We make the approximation that the consistency between richness bins is sufficient to fit the projection component jointly to data from all richness bins.

\section{Colour-magnitude dependence of \redmapper{} photometric-redshift biases}

\begin{figure*}
\includegraphics[width=0.9\textwidth]{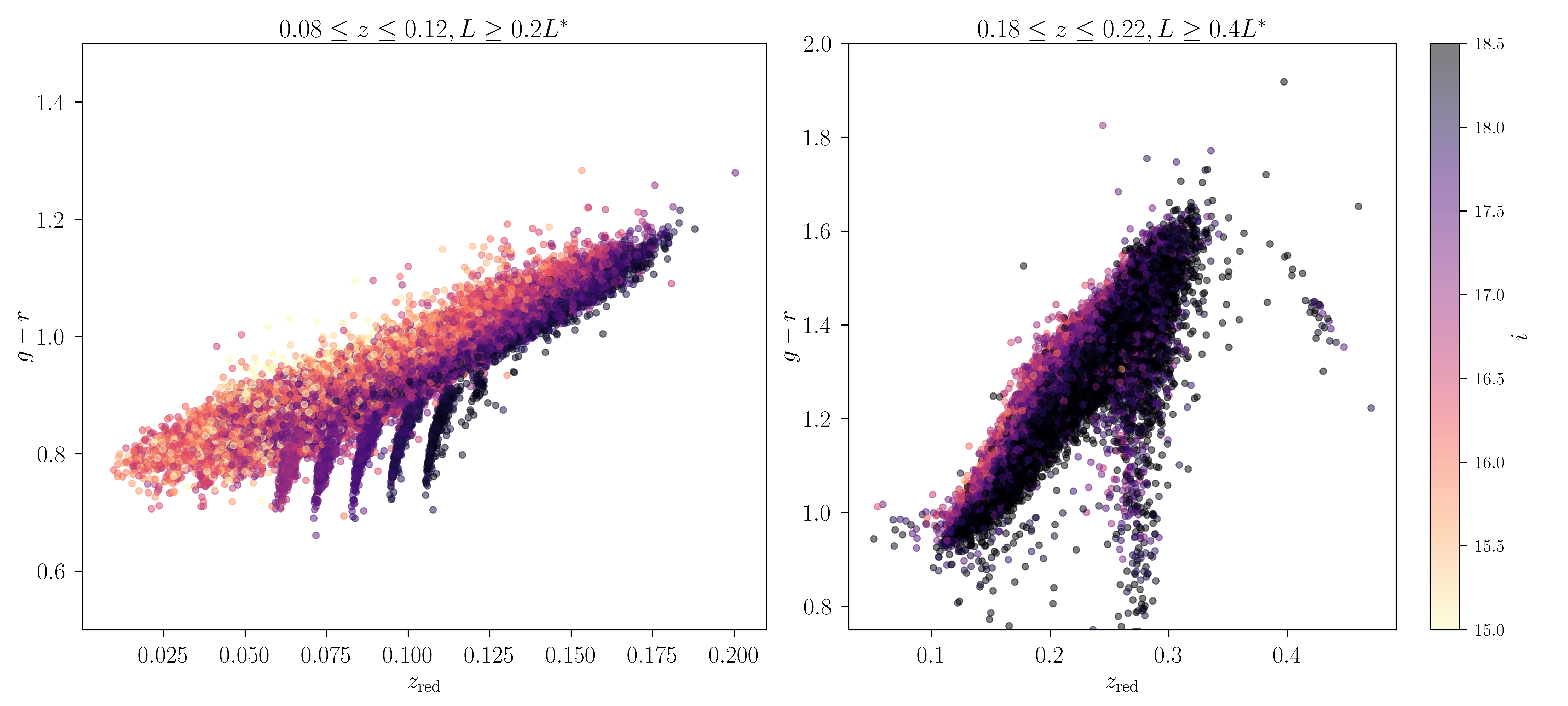}
\caption[Colour vs. \redmapper{} photo-$z$]{Colour-magnitude-dependent photometric redshift failure modes at redshifts $0.1$ (left) and $0.2$ (right). Shown are \redmapper{} cluster galaxy $g-r$ colour as a function of $z_{\mathrm{red}}$, the redshift maximizing the likelihood of a galaxy's photometry being consistent with the \redmapper{} red-sequence. $z_{\mathrm{red}}$ is a kind of photometric redshift that assumes the galaxies are sufficiently consistent with the \redmapper{} red sequence. Streaks in this distribution demonstrate failure modes of the algorithm, justifying re-training the \redmapper{} red-sequence with DESI data. More specifically, this distribution illustrates how generally $g-r$ colour determines photometric redshift at these redshifts, due to the presence of the 4000 \AA{} break in the $g$ band at $z\lesssim0.2$ ($\gtrsim80$\% transmission from roughly 4100 to 5100 \AA{} and much greater transmission than $u$ at 4000 \AA{}). Streaks in this distribution correspond to galaxies with too low $g-r$ to be truly good fits to the red-sequence, i.e., galaxies which are too blue. These galaxies are overwhelmingly drawn from the faint $i$-band magnitude end. This indicates that faint galaxies can disproportionately drive projection effects.
\label{fig:bias-photoz-dep-color}}
\end{figure*}

\begin{figure*}
\includegraphics[width=0.9\textwidth]{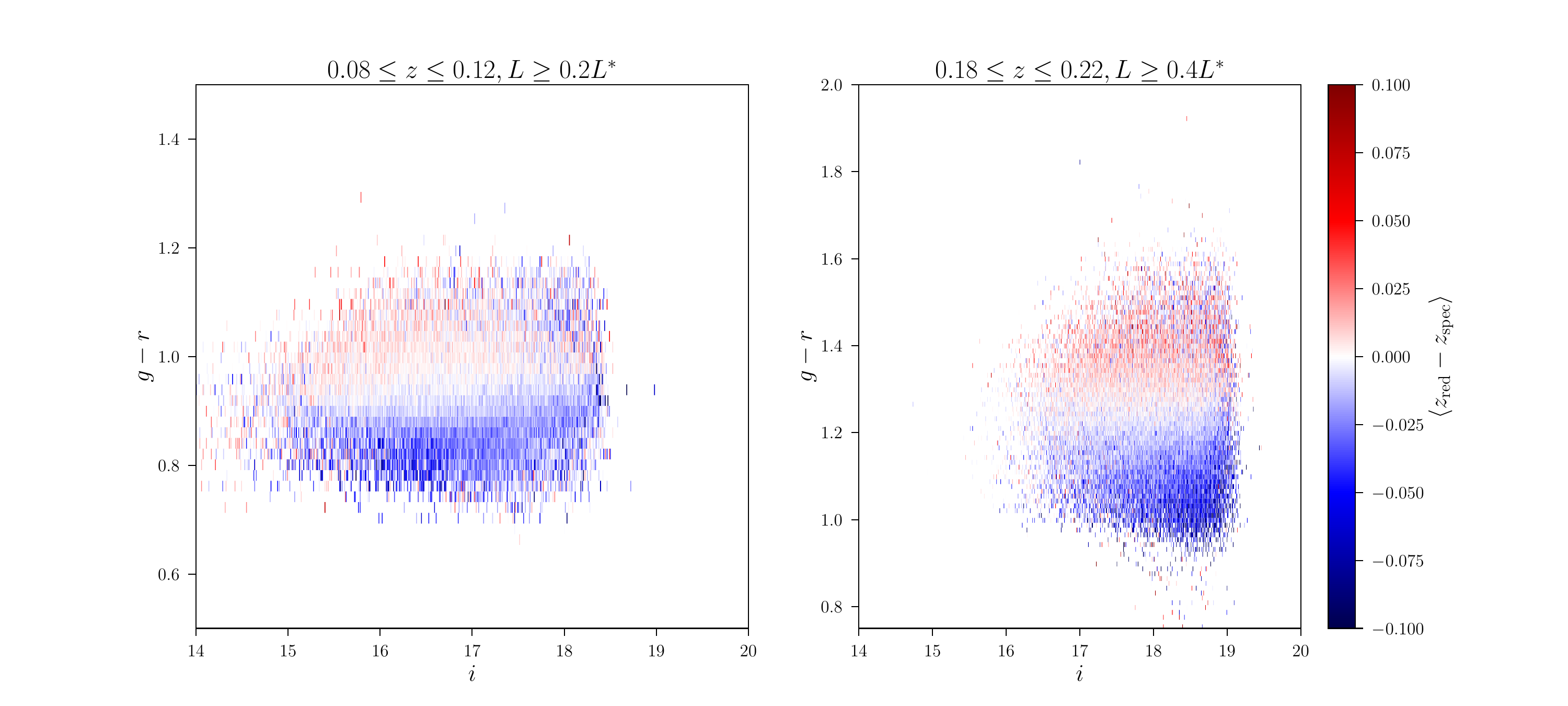}
\caption[Photo-$z$ bias in colour-magnitude bins]{Colour-magnitude dependence of photometric redshift biases at redshifts 0.1 (left) and 0.2 (right). Shown is the mean bias of $z_{\text{red}}$ of redMaPPer cluster galaxies in bins of $g-r$ colour and $i$-band magnitude. 
Overall we see at the red end of the red sequence, the photo-zs are high with respect to $z_{\text{spec}}$ on average. By contrast, at the blue-galaxy end, the photo-zs are low with respect to $z_{\text{spec}}$ on average. For colors near the middle of the distribution, the photo-zs are most accurate. The spectroscopy therefore shows that it may be possible to construct a more robust richness with additional tuning of the red-sequence model scatter or as an outcome of re-training the red sequence model with additional spectroscopic data.}
\label{fig:bias-photoz-dep-color-mag}
\end{figure*}

Projection effects are inherent to optical cluster finding due to irreducible photometric redshift uncertainty. In light of this, we expect projection effects to exhibit colour-magnitude dependence. In this study, we highlight a colour-magnitude dependence in photometric redshift failure. The colour and magnitude dependence of photometric redshift failure and bias is illustrated by Figures \ref{fig:bias-photoz-dep-color} and \ref{fig:bias-photoz-dep-color-mag}. We identify streaks in distribution of $g-r$ colour vs. \redmapper{} galaxy photometric redshift for galaxies associated with clusters in the two redshift ranges analysed in this work. At $z\sim0.1$, the streaks shown emerge from the smoother red-sequence overall when sub-selecting a relatively narrow range of cluster photometric redshift. These streaks deviating from the red sequence are driven by the faint galaxies and reflect photo-$z$ failures. 

This result may reflect the limitation in spectroscopic red-sequence model training data available at the time. Re-training the \redmapper{} red-sequence model with DESI data, as discussed further in \S \ref{sec:discussion}, may mitigate this problem. We emphasize that this result is specific to the SDSS DR8 \redmapper{} catalogue, and that catalogues made afterward have different red-sequence training data. Nevertheless, this issue may persist in general with future use of the algorithm due to spectroscopic training data being insufficiently deep relative to the imaging data. 

\section*{Affiliations}
\label{sec:affiliations}

\vspace{0.4cm}
$^{1}$ Department of Astrophysical Sciences, Princeton University, Princeton NJ 08544, USA\\
$^{2}$ Excellence Cluster ORIGINS, Boltzmannstrasse 2, D-85748 Garching, Germany\\
$^{3}$ University Observatory, Faculty of Physics, Ludwig-Maximilians-Universit\"{a}t, Scheinerstr. 1, 81677 M\"{u}nchen, Germany\\
$^{4}$ Department of Astronomy and Astrophysics, UCO/Lick Observatory, University of California, 1156 High Street, Santa Cruz, CA 95064, USA\\
$^{5}$ Physics Department, Stanford University, Stanford, CA 93405, USA\\
$^{6}$ SLAC National Accelerator Laboratory, 2575 Sand Hill Road, Menlo Park, CA 94025, USA\\
$^{7}$ Lawrence Berkeley National Laboratory, 1 Cyclotron Road, Berkeley, CA 94720, USA\\
$^{8}$ Department of Physics, Boston University, 590 Commonwealth Avenue, Boston, MA 02215 USA\\
$^{9}$ Dipartimento di Fisica ``Aldo Pontremoli'', Universit\`a degli Studi di Milano, Via Celoria 16, I-20133 Milano, Italy\\
$^{10}$ INAF-Osservatorio Astronomico di Brera, Via Brera 28, 20122 Milano, Italy\\
$^{11}$ Department of Physics \& Astronomy, University College London, Gower Street, London, WC1E 6BT, UK\\
$^{12}$ Institut d'Estudis Espacials de Catalunya (IEEC), c/ Esteve Terradas 1, Edifici RDIT, Campus PMT-UPC, 08860 Castelldefels, Spain\\
$^{13}$ Institute of Space Sciences, ICE-CSIC, Campus UAB, Carrer de Can Magrans s/n, 08913 Bellaterra, Barcelona, Spain\\
$^{14}$ Instituto de F\'{\i}sica, Universidad Nacional Aut\'{o}noma de M\'{e}xico,  Circuito de la Investigaci\'{o}n Cient\'{\i}fica, Ciudad Universitaria, Cd. de M\'{e}xico  C.~P.~04510,  M\'{e}xico\\
$^{15}$ NSF NOIRLab, 950 N. Cherry Ave., Tucson, AZ 85719, USA\\
$^{16}$ University of California, Berkeley, 110 Sproul Hall \#5800 Berkeley, CA 94720, USA\\
$^{17}$ Departamento de F\'isica, Universidad de los Andes, Cra. 1 No. 18A-10, Edificio Ip, CP 111711, Bogot\'a, Colombia\\
$^{18}$ Observatorio Astron\'omico, Universidad de los Andes, Cra. 1 No. 18A-10, Edificio H, CP 111711 Bogot\'a, Colombia\\
$^{19}$ Institute of Cosmology and Gravitation, University of Portsmouth, Dennis Sciama Building, Portsmouth, PO1 3FX, UK\\
$^{20}$ University of Virginia, Department of Astronomy, Charlottesville, VA 22904, USA\\
$^{21}$ Fermi National Accelerator Laboratory, PO Box 500, Batavia, IL 60510, USA\\
$^{22}$ Center for Cosmology and AstroParticle Physics, The Ohio State University, 191 West Woodruff Avenue, Columbus, OH 43210, USA\\
$^{23}$ Department of Physics, The Ohio State University, 191 West Woodruff Avenue, Columbus, OH 43210, USA\\
$^{24}$ The Ohio State University, Columbus, 43210 OH, USA\\
$^{25}$ Department of Physics, The University of Texas at Dallas, 800 W. Campbell Rd., Richardson, TX 75080, USA\\
$^{26}$ Department of Physics, Southern Methodist University, 3215 Daniel Avenue, Dallas, TX 75275, USA\\
$^{27}$ Department of Physics and Astronomy, University of California, Irvine, 92697, USA\\
$^{28}$ Sorbonne Universit\'{e}, CNRS/IN2P3, Laboratoire de Physique Nucl\'{e}aire et de Hautes Energies (LPNHE), FR-75005 Paris, France\\
$^{29}$ Departament de F\'{i}sica, Serra H\'{u}nter, Universitat Aut\`{o}noma de Barcelona, 08193 Bellaterra (Barcelona), Spain\\
$^{30}$ Institut de F\'{i}sica d’Altes Energies (IFAE), The Barcelona Institute of Science and Technology, Edifici Cn, Campus UAB, 08193, Bellaterra (Barcelona), Spain\\
$^{31}$ Instituci\'{o} Catalana de Recerca i Estudis Avan\c{c}ats, Passeig de Llu\'{\i}s Companys, 23, 08010 Barcelona, Spain\\
$^{32}$ Department of Physics and Astronomy, Siena College, 515 Loudon Road, Loudonville, NY 12211, USA\\
$^{33}$ Department of Physics \& Astronomy and Pittsburgh Particle Physics, Astrophysics, and Cosmology Center (PITT PACC), University of Pittsburgh, 3941 O'Hara Street, Pittsburgh, PA 15260, USA\\
$^{34}$ IRFU, CEA, Universit\'{e} Paris-Saclay, F-91191 Gif-sur-Yvette, France\\
$^{35}$ Instituto de Astrof\'{i}sica de Andaluc\'{i}a (CSIC), Glorieta de la Astronom\'{i}a, s/n, E-18008 Granada, Spain\\
$^{36}$ Departament de F\'isica, EEBE, Universitat Polit\`ecnica de Catalunya, c/Eduard Maristany 10, 08930 Barcelona, Spain\\
$^{37}$ Department of Physics and Astronomy, Sejong University, 209 Neungdong-ro, Gwangjin-gu, Seoul 05006, Republic of Korea\\
$^{38}$ CIEMAT, Avenida Complutense 40, E-28040 Madrid, Spain\\
$^{39}$ Department of Physics, University of Michigan, 450 Church Street, Ann Arbor, MI 48109, USA\\
$^{40}$ University of Michigan, 500 S. State Street, Ann Arbor, MI 48109, USA\\


\bsp	
\label{lastpage}
\end{document}